# Generalized coherent states with shifted (displaced) arguments [1]

**Dušan POPOV**[a, b]

[a] University Politehnica Timisoara, Department of Physical Foundations of Engineering, B-dul Vasile Pârvan No. 2, 300223 Timisoara, Romania
[b] Serbian Academy of Nonlinear Sciences (SANS), Kneza Mihaila 36, Beograd-Stari Grad, Belgrade, Serbia
E-mail: dusan_popov@yahoo.co.uk
ORCID: https://orcid.org/0000-0003-3631-3247

## Abstract

In the paper we developed a procedure for constructing generalized coherent states with shifted argument, as a result of the action of the generalized displacement operator. This was based on the action of a pair of nonlinear ladder operators, which generate nonlinear coherent states. To examine the properties of coherent states with shifted argument, the rules of the normal operator ordering technique (DOOT) were used. The results obtained were verified for a series of particular cases, obtaining expressions consistent with those in the literature. The expressions obtained will be, among others, useful in quantum optics, e.g. for calculating the Wigner operator in the representation of coherent states.

**Key words:** coherent states; displacement operator; hypergeometric function.

## 1. Introduction

It is a well-known fact that in many branches of physics the so-called displacement operator is used. In essence, this operator performs a displacement or a translation of the variable $x$ characteristic of a certain physical quantity, in the same space, but with the variable displaced by $\Delta x$. It results the same function but depending on the shifted variable $x + \Delta x$. Generally, the unitary shifted or displacement operator is an exponential function $\hat{\mathcal{D}}(x) = \exp\left( \Delta x \frac{\partial}{\partial x} \right)$ and has the role of transferring an initial state or function, e.g. $\Phi(x)$, which depends on some argument $x$, to another state or function, with a shifted argument $\Phi(x+\Delta x)$:

$$\exp\left(\Delta x \frac{\partial}{\partial x}\right)\Phi(x) = \Phi(x+\Delta x) \tag{1.01}$$

[1] The article is intended to be a tribute to **Emeritus Professor Hong-Yi FAN**, from University of Science and Technology of China (USTC), China, for his contribution to the creation and promotion of the IWOP calculation technique in Quantum Mechanics.

In quantum mechanics the displacement operators are used to change the position $q$ and / or the moment $p$ in the phase space. Depending on these variables, as well as on the associated generators $\hat{\mathcal{Q}}$ and $\hat{\mathcal{P}}$, with $\left[\hat{\mathcal{Q}}\,,\,\hat{\mathcal{P}}\right] = \mathrm{i}$, and $\hbar = 1$, the displacement operator is

$$\hat{\mathcal{D}}(x,p) = \exp\left(\mathrm{i}\,p\,\hat{\mathcal{Q}} - \mathrm{i}\,q\,\hat{\mathcal{P}}\right) \tag{1.02}$$

The next step, towards the transition to quantum optics, is the introduction of the complex variable $z$, as well as to the canonical creation $\hat{\boldsymbol{a}}^{+}$ and annihilation $\hat{\boldsymbol{a}}$ operators, $\left[\hat{\boldsymbol{a}}\,,\,\hat{\boldsymbol{a}}^{+}\right] = 1$ :

$$z = \frac{1}{\sqrt{2}}(q + \mathrm{i}\,p) \;,\quad z^{*} = \frac{1}{\sqrt{2}}(q - \mathrm{i}\,p) \;,\quad \hat{\boldsymbol{a}} = \frac{1}{\sqrt{2}}\left(\hat{\mathcal{Q}} + \mathrm{i}\hat{\mathcal{P}}\right) \;,\quad \hat{\boldsymbol{a}}^{+} = \frac{1}{\sqrt{2}}\left(\hat{\mathcal{Q}} - \mathrm{i}\hat{\mathcal{P}}\right) \tag{1.03}$$

At the same time, the inverse relationships are

$$q = \frac{1}{\sqrt{2}}\left(z^{*} + z\right) \;,\quad p = \frac{\mathrm{i}}{\sqrt{2}}\left(z^{*} - z\right) \;,\quad \hat{\mathcal{Q}} = \frac{1}{\sqrt{2}}\left(\hat{\boldsymbol{a}}^{+} + \hat{\boldsymbol{a}}\right) \;,\quad \hat{\mathcal{P}} = \frac{\mathrm{i}}{\sqrt{2}}\left(\hat{\boldsymbol{a}}^{+} - \hat{\boldsymbol{a}}\right) \tag{1.04}$$

Since the complex variable $z$ intervenes in the formalism of CSs (CSs), using the above relations it is obtained that in quantum optics the displacement operator will have the expression

$$\hat{\mathcal{D}}(z) = \exp\left(z\,\hat{\boldsymbol{a}}^{+} - z^{*}\hat{\boldsymbol{a}}\right) \tag{1.05}$$

But in the real physical world there are phenomena whose description requires the use of more complicated operators than the canonical ones, corresponding to the linear or one-dimensional harmonic oscillator. If the canonical operators generate linear CSs (LCSs), the other more complicated operators correspond to nonlinear CSs (NCSs). Therefore, it is not difficult to assume that within the formalism of nonlinear or generalized CSs (NCSs), the generalized displacement operator will be used.

The formalism of LCSs achieved a wide development with the introduction of the technique of integration within an ordered product (IWOP). This real "revolution" of calculation, which makes full use of Dirac's symbolic method of representation, was carried out by Hong-yi Fan and collaborators, since 1982 (see, [1], [2], [3], [4] and references therein), making the calculations related to the formalism of CSs become more transparent and easier to interpret. Unfortunately, the authors and users of the IWOP technique only focused on canonical CSs, without trying to generalize this technique to nonlinear CSs as well.

In an attempt to generalize the IWOP technique and make it applicable to non-linear systems, other than those of the one-dimensional oscillator, we developed a similar technique that we called *diagonal ordering operation technique* (DOOT) [5], which we successfully applied for different nonlinear oscillators, characterized by different kinds of CSs [6], [7], [8], [9], [10].

Consequently, the construction of the generalized displacement operator $\hat{\mathcal{D}}(z)$ is based on the *DOOT - IWOP correspondence principle*, which is part of a broader framework of the correspondence principle between the characteristic quantities of a nonlinear system versus those of a linear system. The DOOT - IWOP correspondence principle assumes that each characteristic quantity within the DOOT technique, $X_{\text{DOOT}}$, corresponds to a characteristic quantity within the IWOP technique, $X_{\text{IWOP}}$, according to the following limit:

$$\lim_{\substack{p=q \\ \tilde{\alpha}=\tilde{b}}} X_{\text{DOOT}} = X_{\text{IWOP}} \quad , \quad \lim_{\substack{p=q \\ \tilde{\alpha}=\tilde{b}}} \hat{\mathcal{D}}(z) = \hat{\mathbf{D}}(z) \tag{1.06}$$

where, as will be seen below, the integers $p$ and $q$, respectively the sets of real numbers $\tilde{\alpha}$ and $\tilde{b}$ are characteristic of the DOOT technique.

In this article, we aim to apply the rules of the DOOT technique to examine the expressions and properties of CSs with shifted or displaced arguments. All the partial results obtained were verified (tested), applying the above limit, for the case of canonical CSs, characteristic of the one-dimensional harmonic oscillator (HO-1D). In this sense, the article can be considered as a tribute to Professor Hong-Yi Fan, one of the initiators and a fervent promoter of the IWOP calculation technique, applied to the case of canonical CSs. By applying this technique, not only a series of issues related to quantum operators, related to CSs, became clearer, but also a series of new results / formulas useful in the mathematics of complex functions were obtained.

## 2. The case of canonical (linear) coherent states

In this section we will reiterate a series of notions related to canonical CSs, as well as the displacement operators, which will be useful to us further. It is well-known that for the one-dimensional harmonic oscillator (HO-1D) the normalized (or canonical) CSs have the same expression for all three ways of definition (Barut-Girardello – BG-CSs, Klauder-Perelomov – KP-CSs, and Gazeau-Klauder – GK-CSs) [11]. These states can be expanded in an orthogonal basis, e.g. the basis of Fock vector's, i.e. the eigenstates $|n>$ of the number operator $\hat{\mathcal{N}}\,|n>=n|n>$. We consider the non-negative integers which characterize the energy spectrum of the one-dimensional quantum harmonic oscillator $\left\{|n>,\ <n|m>=\delta_{nm},\ n,m=0,1,2,...,\infty\right\}$. Then, this development is

$$| z >= e^{-\frac{1}{2}|z|^2} \sum_{n=0}^{\infty} \frac{z^n}{\sqrt{n!}} | n > \qquad (2.01)$$

labeled by the complex variable $z = | z | \exp(\mathrm{i}\varphi),\ 0 \le | z | \le \infty\ ,\ 0 \le \varphi \le 2\pi$ .

This expression can also be obtained with the help of the displacement operator

$$\hat{\mathcal{D}}(z) = e^{z\hat{a}^+ - z^*\hat{a}} = e^{-\frac{1}{2}|z|^2} e^{z\hat{a}^+} e^{-z^*\hat{a}} \qquad (2.02)$$

obtained using the Baker–Campbell–Hausdorff formula

$$\exp\left(\hat{\mathcal{X}} + \hat{\mathcal{Y}}\right) = \exp\left(\hat{\mathcal{X}}\right)\exp\left(\hat{\mathcal{Y}}\right)\exp\left(-\frac{1}{2}\left[\hat{\mathcal{X}}, \hat{\mathcal{Y}}\right]\right) \qquad (2.03)$$

As seen, canonical ladder operators (the creation $\hat{a}^+$, respectively annihilation $\hat{a}$) intervene in the expression of the displacement operator. They satisfy the commutation relation $\left[\hat{a}\ ,\hat{a}^+\right] = 1$ and their action on the Fock vectors are

$$\begin{array}{ll}
\hat{a}| n >= \sqrt{n}\, | n-1 > & ,\ < n | \hat{a}^+ = \sqrt{n} < n-1 | \\
\hat{a}^s | n >= \sqrt{\frac{n!}{(n-s)!}}\, | n-s > & ,\ < n | \left(\hat{a}^+\right)^s = < n-s | \sqrt{\frac{n!}{(n-s)!}}\ ,\ 0 \le s \le n \\
\hat{a}^+ | n >= \sqrt{n+1}\, | n+1 > & ,\ < n | \hat{a} = \sqrt{n+1} < n+1 | \\
\left(\hat{a}^+\right)^s | n >= \sqrt{\frac{(n+s)!}{n!}}\, | n+s > & ,\ < n | \hat{a}^s = < n+s | \sqrt{\frac{(n+s)!}{n!}} \\
\hat{a}^n | 0 >= \sqrt{n!}\, | n > & ,\ < 0 | \left(\hat{a}^+\right)^n = \sqrt{n!} < n | \\
< n | \hat{a}^+\hat{a} | n >= n & ,\ < n | \hat{a}\hat{a}^+ | n >= n+1
\end{array} \qquad (2.04)$$

To these operators we can added also the particle number operator (sometimes called only the number operator) $\hat{\mathcal{N}} \equiv \hat{a}^+\hat{a}$ , which will be useful in the next. This is a self-adjoint operator, i.e. it is equal to its Hermitian conjugate (or complex conjugate transpose), $\hat{\mathcal{N}} = \hat{\mathcal{N}}^+$.

The CSs can be obtained by applying the displacement operator $\hat{\mathcal{D}}\,(z)$ on the ket vector of the vacuum state $| 0 >$ and taking into account that $\hat{a}\, | 0 >= | 0 >$ , respectively $\exp\left(-z^*\hat{a}\right) | 0 >= | 0 >$:

$$| z >= \hat{\mathcal{D}}\,(z) | 0 >= e^{-\frac{1}{2}|z|^2} e^{z\hat{a}^+} e^{-z^*\hat{a}} | 0 > \qquad (2.05)$$

Their counterpart, i.e. their hermitian conjugate is

$$< 0 | \hat{\mathcal{D}}^+(z) = < 0 | \hat{\mathcal{D}}(-z) = e^{z^*\hat{a} - z\hat{a}^+} = e^{\frac{1}{2}|z|^2} e^{z^*\hat{a}} e^{-z\hat{a}^+} \qquad (2.06)$$

so that

$$< z | = < 0 | \hat{\mathcal{D}}^{+}(z) \tag{2.07}$$

Since CSs are normalized to unity

$$< z | = < 0 | \hat{\mathcal{D}}^{+}(z) \hat{\mathcal{D}}(z) | 0 > = 1 \tag{2.08}$$

it follows that the displacement operator is a unitary operator

$$\hat{\mathcal{D}}^{+}(z) \hat{\mathcal{D}}(z) = \hat{\mathcal{D}}(z) \hat{\mathcal{D}}^{+}(z) = 1 \tag{2.09}$$

If we apply the displacement operator (which depends on another variable) on the previously obtained CSs, we will have, successively:

$$\hat{\mathcal{D}}(\sigma) | z > = e^{-\frac{1}{2}|\sigma|^2} e^{\sigma \hat{a}^{+}} | z > = e^{-\frac{1}{2}|\sigma|^2} e^{-\frac{1}{2}|z|^2} \sum_{l=0}^{\infty} \frac{z^l}{\sqrt{l!}} e^{\sigma \hat{a}^{+}} | l > \tag{2.10}$$

$$e^{\sigma \hat{a}^{+}} | l > = \sum_{m=0}^{\infty} \frac{\sigma^m}{m!} \left(\hat{a}^{+}\right)^m | l > = \sum_{m=0}^{\infty} \frac{\sigma^m}{m!} \sqrt{\frac{(l+m)!}{l!}} | l+m > \tag{2.11}$$

$$\begin{aligned} \hat{\mathcal{D}}(\sigma) | z > &= e^{-\frac{1}{2}|\sigma|^2} e^{-\frac{1}{2}|z|^2} \sum_{l=0}^{\infty} \frac{z^l}{\sqrt{l!}} \sum_{m=0}^{\infty} \frac{\sigma^m}{m!} \sqrt{\frac{(l+m)!}{l!}} | l+m > = \\ &= \sum_{m=0}^{\infty} \sum_{l=0}^{\infty} \frac{1}{\sqrt{(l+m)!}} \frac{(l+m)!}{l! m!} z^l \sigma^m | l+m > \end{aligned} \tag{2.12}$$

If we make the substitution $l+m=n, \;\; l=n-m \geq 0, \;\; l \leq n$

$$\begin{aligned} \hat{\mathcal{D}}(\sigma) | z > &= e^{-\frac{1}{2}|\sigma|^2} e^{-\frac{1}{2}|z|^2} \sum_{n-l=0}^{\infty} \frac{1}{\sqrt{n!}} \sum_{m=0}^{n} \frac{n!}{(n-m)!m!} z^m \sigma^{n-m} | n > = \\ &= e^{-\frac{1}{2}|\sigma|^2} e^{-\frac{1}{2}|z|^2} \sum_{n=0}^{\infty} \frac{1}{\sqrt{n!}} \left[ \sum_{m=0}^{n} \binom{n}{m} z^m \sigma^{n-m} \right] | n > = \\ &= e^{-\frac{1}{2}|\sigma|^2} e^{-\frac{1}{2}|z|^2} \sum_{n=0}^{\infty} \frac{(z+\sigma)^n}{\sqrt{n!}} | n > = \frac{e^{-\frac{1}{2}|\sigma|^2} e^{-\frac{1}{2}|z|^2}}{e^{-\frac{1}{2}|z+\sigma|^2}} | z+\sigma > \end{aligned} \tag{2.13}$$

Because the right bracket no longer depends on l, it has been removed from the first sum (over the index *n*). Consequently, CSs with shifted arguments are

$$| z+\sigma > = e^{-\frac{1}{2}\left(\sigma^* z + \sigma z^*\right)} \hat{\mathcal{D}}(\sigma) \hat{\mathcal{D}}(z) | 0 > = \hat{\mathcal{D}}(\sigma + z) | 0 > \tag{2.14}$$

which means that the successive displacements are equivalent to a single displacement up to an phase factor. The above relation can be written in an equivalent manner,

$$| z+\sigma >= e^{-\frac{1}{2}|z+\sigma|^2} \sum_{n=0}^{\infty} \frac{(z+\sigma)^n}{\sqrt{n!}} | n > \tag{2.15}$$

This expression of CSs with shifted argument could also be deduced directly, if we consider that the sum of two complex numbers $z$ and $\sigma$ is also a complex number $z+\sigma$. It is observed that the expression on the right-hand side is symmetric in the variables $z$ and $\sigma$, which thus have "equal rights".

## 3. Generalized coherent states

For several decades since the introduction, in 1926, by Schrodinger, of the notion of coherent state [12], the scientific world was only interested in the canonical CSs, connected with the one-dimensional harmonic oscillator. Over time, however, other types of CSs began to appear and were named nonlinear CSs (NCSs) or, more recently, generalized CSs (GCSs). As a counterpart, the canonical CSs of HO-1D were named linear CSs.

Mainly, any nonlinear or generalized CSs have the following structure:

$$| z >= \frac{1}{\sqrt{N(| z |^2)}} \sum_{n=0}^{\infty} \frac{z^n}{\sqrt{\rho(n)}} | n >= \frac{1}{\sqrt{N(| z |^2)}} N(z\hat{\mathcal{A}}_+) | 0 > \tag{3.01}$$

where $\rho(n)$ are positive numbers, called *structure constants*, because they determine the structure and character of the normalization function. This must be an analytical function:

$$N(| z |^2) = \sum_{n=0}^{\infty} \frac{1}{\rho(n)} (| z |^2)^n \ . \tag{3.02}$$

In essence, generalized CSs are generated by a hermitian conjugates pair of ladder operators: the creation $\hat{\mathcal{A}}_+$ and the $\hat{\mathcal{A}}_-$ annihilation, where $(\hat{\mathcal{A}}_+)^+ = \hat{\mathcal{A}}_-$ and $(\hat{\mathcal{A}}_-)^+ = \hat{\mathcal{A}}_+$. These operators are connected with the canonical operators $\hat{a}^+$ and $\hat{a}$ through a function dependent of the particle number operator $\hat{\mathcal{N}} = \hat{a}^+\hat{a}$. This real function $f(\hat{\mathcal{N}})$ is called the nonlinearity function.

Nonlinear operators have the following structure

$$\hat{\mathcal{A}}_- = \hat{a}\, f(\hat{\mathcal{N}}) \ , \quad \hat{\mathcal{A}}_+ = f(\hat{\mathcal{N}})\hat{a}^+ \tag{3.03}$$

The nonlinearity function $f(\hat{\mathcal{N}})$ can be, up to a point, arbitrary, as long as it ensures that the normalization function is an analytic function (non-analytic ones having no physical meaning). Moreover, the nonlinearity function $f(\hat{\mathcal{N}})$ satisfy the following general relation

$$f(\hat{\mathcal{N}}+m) | n+p >= f(n+p+m) | n+p > \tag{3.04}$$

At the limit the pair of newly introduced operators reduces to the canonical operators:

$$\lim_{\substack{p=q\\ \boldsymbol{a}=\boldsymbol{b}}} \hat{\mathcal{A}}_- = \lim_{f(\hat{N})\to 1} \hat{\mathcal{A}}_- = \hat{\boldsymbol{a}} \quad , \quad \lim_{\substack{p=q\\ \boldsymbol{a}=\boldsymbol{b}}} \hat{\mathcal{A}}_+ = \lim_{f(\hat{N})\to 1} \hat{\mathcal{A}}_+ = \hat{\boldsymbol{a}}^+ \tag{3.05}$$

We will choose the nonlinearity function in the following manner

$$f(\hat{\mathcal{N}}) = \sqrt{\frac{\prod_{j=1}^{q}\left(b_j - 1 + \hat{\mathcal{N}}\right)}{\prod_{i=1}^{p}\left(a_i - 1 + \hat{\mathcal{N}}\right)}} \quad , \qquad f(\hat{\mathcal{N}})\,|\,n> = \sqrt{\frac{\prod_{j=1}^{q}\left(b_j - 1 + n\right)}{\prod_{i=1}^{p}\left(a_i - 1 + n\right)}}\,|\,n> \tag{3.06}$$

The actions of these operators and their hermitian conjugates on the Fock vectors is

$$\begin{gathered} \hat{\mathcal{A}}_-\,|\,n> = \sqrt{e(n)}\,|\,n-1> \quad , \quad <n\,|\,\hat{\mathcal{A}}_+ = \sqrt{e(n)} < n-1\,| \ , \\ \hat{\mathcal{A}}_+\,|\,n> = \sqrt{e(n+1)}\,|\,n+1> \ , \quad <n\,|\,\hat{\mathcal{A}}_- = \sqrt{e(n+1)} < n+1\,| \\ <n\,|\,\hat{\mathcal{A}}_+\hat{\mathcal{A}}_-\,|\,n> = e(n) \end{gathered} \tag{3.07}$$

Here we used the notation:

$$e(n) \equiv n\left[f(n)\right]^2 = n\frac{\prod_{j=1}^{q}\left(b_j - 1 + n\right)}{\prod_{i=1}^{p}\left(a_i - 1 + n\right)} \tag{3.08}$$

The repeated action of the creation operator on the vacuum ket vector $|\,0>$ (and similarly for their hermitian conjugate $<0\,|$) is summarized as follows:

$$\begin{gathered} \left(\hat{\mathcal{A}}_+\right)^n |\,0> = \sqrt{\prod_{s=1}^{n} e(s)}\,|\,n> \equiv \sqrt{\rho(n)}\,|\,n> \\ <0\,|\left(\hat{\mathcal{A}}_-\right)^n = <n\,|\sqrt{\prod_{s=1}^{n} e(s)} \equiv <n\,|\sqrt{\rho(n)} \end{gathered} \tag{3.09}$$

where we introduced a new notation

$$\rho(n) = \prod_{s=1}^{n} e(s) = n!\left(\left[f(n)\right]!\right)^2 = n!\frac{\prod_{j=1}^{q}\left(b_j\right)_n}{\prod_{i=1}^{p}\left(a_i\right)_n} \quad , \quad \lim_{f(\hat{N})\to 1}\rho(n) = n! \quad , \quad \rho(0) = 1 \tag{3.10}$$

Anticipating what follows, let us point out that the quantity $\rho(n)$ is called the *structure function*, because it determines the internal structure of CSs.

Here we used the *functional factorial* defined as $[x(n)]! \equiv x(1)x(2)...x(n)$, $[x(0)]! = 1$ as well as the Pochhammer symbol $(x)_n = \prod_{s=1}^{n}(x+s-1) = \Gamma(x+n) \,/\, \Gamma(x)$. Consequently, the structure function can be written as

$$\rho(n) = n! \frac{\prod_{j=1}^{q}(b_j)_n}{\prod_{i=1}^{p}(a_i)_n} \equiv n![g(n)]! \quad , \quad [g(n)]! \equiv \frac{\prod_{j=1}^{q}(b_j)_n}{\prod_{i=1}^{p}(a_i)_n} \quad , \quad [g(0)]! = 1 \tag{3.11}$$

This entity, in which a function that depends on a number appears instead of a number, can be called *functional factorial*.

In the next we will also use the following short notations:

$$\begin{pmatrix}\rho(l)\\ \rho(n)\end{pmatrix} = \frac{\rho(l)}{\rho(n)\,\rho(l-n)} \quad , \quad \begin{pmatrix}[g(l)]\\ [g(n)]\end{pmatrix} \equiv \frac{[g(l)]!}{[g(n)]!\,[g(l-n)]!}$$
$$\begin{pmatrix}\rho(l)\\ \rho(n)\end{pmatrix} = \begin{pmatrix}l\\ n\end{pmatrix}\begin{pmatrix}[g(l)]\\ [g(n)]\end{pmatrix} \tag{3.12}$$

The entity $\begin{pmatrix}\rho(l)\\ \rho(n)\end{pmatrix}$ can be interpreted as the *functional binomial coefficients*, because at the limit, this turns into the usual binomial coefficient:

$$\lim_{\tilde{a}=\tilde{b}=1}\begin{pmatrix}\rho(l)\\ \rho(n)\end{pmatrix} = \begin{pmatrix}l\\ n\end{pmatrix} \tag{3.13}$$

Generally, the ladder operators $\hat{\mathcal{A}}_+$ and $\hat{\mathcal{A}}_-$ are not commutable, but, as we will see below, they obey the rules of the *Diagonal Operator Ordering Technique* (DOOT), which we introduced in a previous paper [5]. The main rules of the DOOT are:

a) Inside the symbol #...# the order of operators $\hat{\mathcal{A}}_+$ and $\hat{\mathcal{A}}_-$ can be permuted, i. e. $\#(\hat{\mathcal{A}}_-)^m(\hat{\mathcal{A}}_+)^n\# = \#(\hat{\mathcal{A}}_+)^n(\hat{\mathcal{A}}_-)^m\#$, so that finally we obtain a function or an expression that is normally ordered, meaning that operator $\hat{\mathcal{A}}_+$ is located on the left side and operator $\hat{\mathcal{A}}_-$ is on its right side. Consequently, only inside the #...# symbol the ladder operators $\hat{\mathcal{A}}_+$ and $\hat{\mathcal{A}}_-$ commute with each other within ordered product, i.e. $\#\hat{\mathcal{A}}_-\hat{\mathcal{A}}_+\# = \#\hat{\mathcal{A}}_+\hat{\mathcal{A}}_-\#$.

b) Inside the symbol $\#...\#$ we can perform all algebraic operations (derivations, integrations), according to the usual rules.

c) The ladder operators $\hat{\mathcal{A}}_+$ and $\hat{\mathcal{A}}_-$ can be treated as simple c-numbers; d) the vacuum state projector $|0\rangle\langle 0|$, in the frame of DOOT, has the following normal ordered form:

$$|0\rangle\langle 0| = \frac{1}{\sum_{n=0}^{\infty}\frac{1}{\rho(n)}\#\left(\hat{\mathcal{A}}_+\hat{\mathcal{A}}_-\right)^n\#} = \frac{1}{\#\,{}_pF_q\left(\tilde{a}\,;\tilde{b}\,;\hat{\mathcal{A}}_+\hat{\mathcal{A}}_-\right)\#} \tag{3.14}$$

where we used the abbreviated notation for the coefficients: $\tilde{\boldsymbol{a}} \equiv \{a_1, a_2, ..., a_p\} \equiv \{a_i\}_1^p$ and so on, and ${}_pF_q(\tilde{\boldsymbol{a}}\,;\tilde{\boldsymbol{b}}\,;x)$ is the generalized hypergeometric function.

To simplify writing formulas, everywhere possible, without causing confusion, in the next we will write the hypergeometric functions without mentioning the set of numbers $\tilde{\boldsymbol{a}}$ and $\tilde{\boldsymbol{b}}$ and indicating only their argument, e.g. ${}_pF_q(\tilde{\boldsymbol{a}}\,;\tilde{\boldsymbol{b}}\,;x) \equiv {}_pF_q(x)$. Instead, we will use the same $|z>$ notation for both generalized CSs and canonical CSs.

With this notation, the generalized hypergeometric function will be written as

$$ {}_pF_q(\tilde{\boldsymbol{a}}\,;\tilde{\boldsymbol{b}}\,;x) \equiv {}_pF_q(x) = \sum_{n=0}^{\infty} \frac{\prod_{i=1}^{p}(a_i)_n}{\prod_{j=1}^{q}(b_j)_n} \frac{x^n}{n!} = \sum_{n=0}^{\infty} \frac{x^n}{\rho(n)} \tag{3.15} $$

One of the important advantages of using the DOOT technique (as well as the IWOP technique) is that it allows us to express various entities in terms of creation / annihilation operators $\hat{\mathcal{A}}_+$ and $\hat{\mathcal{A}}_-$. Including the vacuum projector $|0><0|$ which is expressed as a function that depends on the normal ordered product of creation and annihilation operators $\#\hat{\mathcal{A}}_+\hat{\mathcal{A}}_-\#$. Thus we end up working with functions that have operators as arguments, and these are viewed as simple *c*-numbers.

Let us emphasize here that the DOOT technique is a generalization of the technique of *Integration Within an Ordered Product* (IWOP), introduced by Hong-yi Fan (see, e.g. [3] and references therein). But, the difference between the two techniques consists in the fact that the IWOP is applicable only for Bose operators, referring to the CSs of the HO-1D, while DOOT applies to any pair of creation / annihilation operators, adopting the above rules. In this sense, the IWOP appears as a particular case of DOOT.

The choice of ladder operators $\hat{\mathcal{A}}_+$ and $\hat{\mathcal{A}}_-$, having the above properties and who obey the rules of the DOOT technique was not accidental. As will be seen below, this choice leads to the fact that the normalization function of CSs will be the generalized hypergeometric function itself, ${}_pF_q(x)$.

Consequently, let us define the CSs in the Barut-Girardello manner, i.e. as the eigenvalues of the annihilation operator

$$ \hat{\mathcal{A}}_- \,|\, z> = z\,|\, z> \quad , \qquad < z\,|\,\hat{\mathcal{A}}_+ = z^* < z\,| \tag{3.16} $$

The second relation can be obtained either by complex conjugation of the first relation, or by using the property $< n\,|\,\hat{\mathcal{A}}_+ = \sqrt{e(n)} < n-1\,|$ , see Eq. (3.07).

Using this definition, as well as the rules of the DOOT technique, it is obtained that the expression of generalized CSs is

$$|z>=\frac{1}{\sqrt{{}_pF_q(|z|^2)}}\sum_{n=0}^{\infty}\frac{z^n}{\sqrt{\rho(n)}}|n>=\frac{1}{\sqrt{{}_pF_q(|z|^2)}}\,{}_pF_q(z\hat{\mathcal{A}}_+)|0> \tag{3.17}$$

and their complex conjugate

$$<z|=\frac{1}{\sqrt{{}_pF_q(|z|^2)}}\sum_{n=0}^{\infty}\frac{(z^*)^n}{\sqrt{\rho(n)}}<n|=\frac{1}{\sqrt{{}_pF_q(|z|^2)}}<0|\,{}_pF_q(z^*\hat{\mathcal{A}}_-) \tag{3.18}$$

Using the DOOT rules, the projector, in the CSs representation can be written as

$$\begin{aligned}|z><z|&=\frac{1}{{}_pF_q(|z|^2)}\#\,{}_pF_q(z\hat{\mathcal{A}}_+)|0><0|\,{}_pF_q(z^*\hat{\mathcal{A}}_-)\#=\\&=\frac{1}{{}_pF_q(|z|^2)}\#\frac{1}{{}_pF_q(\hat{\mathcal{A}}_+\hat{\mathcal{A}}_-)}\,{}_pF_q(z\hat{\mathcal{A}}_+)\,{}_pF_q(z^*\hat{\mathcal{A}}_-)\#\end{aligned} \tag{3.19}$$

Because the normalization function must be an analytical function, their convergence radius $R_c$ of the series must fulfill the inequality: $0<|z|<R_c=\lim_{n\to\infty}\frac{\rho(n)}{\rho(n+1)}$.

In 1963 Klauder established that all generalized CSs (therefore, linear and non-linear) must satisfy the following minimum conditions, which were later called "the Klauder's prescriptions" [13]:

(I). NCSs are normalized but non-orthogonal:

$$<z|z'>=\frac{{}_pF_q(z^*z')}{\sqrt{{}_pF_q(|z|^2)}\sqrt{{}_pF_q(|z'|^2)}}=\begin{cases}1, & z=z'\\ \neq 0, & z\neq z'\end{cases} \tag{3.20}$$

and thus the generalized CSs form an overcomplete set.

(II). Continuity in the label variable $z$ of the NCSs:

$$\lim_{z\to z'}\|z-z'\|=\lim_{\substack{r\to r'\\ \varphi\to\varphi'}}\sqrt{r^2+r'^2-2rr'\cos(\varphi-\varphi')}=0 \tag{3.21}$$

(III). NCSs must satisfy the resolution of the identity, or the closure relation which can be expressed by the resolution of the identity operator $I$ in the vector space of quantum states

$$\int d\mu(z)|z><z|=\hat{I}=\sum_{n=0}^{\infty}|n><n| \tag{3.22}$$

As a consequence of this relation, the property of CSs to form an overcomplete set translates into the fact that any state can be decomposed on the set of CSs:

$$|z'> = \int d\mu(z) |z><z|z'> \tag{3.23}$$

It is not difficult to verify that the integration measure is [5].

$$d\mu(z) = \Gamma\left(\tilde{\boldsymbol{a}} / \tilde{\boldsymbol{b}}\right) \frac{d\varphi}{2\pi} d(|z|^2)\; {}_pF_q(|z|^2)\; G^{q+1,0}_{p,\,q+1}\left( |z|^2 \middle| \begin{matrix} / \,; & \tilde{\boldsymbol{a}}\,\text{-}\,\boldsymbol{1} \\ 0\,, \tilde{\boldsymbol{b}}\,\text{-}\,\boldsymbol{1}\,; & / \end{matrix} \right) \tag{3.24}$$

where we used the notation

$$\Gamma\left(\tilde{\boldsymbol{a}} / \tilde{\boldsymbol{b}}\right) \equiv \frac{\prod_{i=1}^{p} \Gamma(a_i)}{\prod_{j=1}^{q} \Gamma(b_j)} \tag{3.25}$$

Here $G^{q+1,0}_{p,\,q+1}\left(|z|^2 \,|\, ...\right)$ is the Meijer G-function which can be connected with the hypergeometric function ${}_pF_q(\,|z|^2)$ as follows:

$${}_pF_q(|z|^2) = \Gamma\left(\tilde{\boldsymbol{b}} / \tilde{\boldsymbol{a}}\right) G^{1,p}_{p,q+1}\left( -|z|^2 \middle| \begin{matrix} \boldsymbol{1}\,\text{-}\,\tilde{\boldsymbol{a}}\,; & / \\ 0\,; & \boldsymbol{1}\,\text{-}\,\tilde{\boldsymbol{b}} \end{matrix} \right) \tag{3.26}$$

where $\boldsymbol{1}\,\text{-}\,\tilde{\boldsymbol{a}} \equiv \{1\text{-}a_1, 1\text{-}\,a_2, ..., 1\text{-}\,a_p\}$ and so on.

With the above considerations, the relation of decomposition of unity operator can be also written as

$$\int \frac{d^2 z}{\pi} G^{q+1,0}_{p,\,q+1}\left( |z|^2 \middle| \begin{matrix} / \,; & \tilde{\boldsymbol{a}}\,\text{-}\,\boldsymbol{1} \\ 0\,, \tilde{\boldsymbol{b}}\,\text{-}\,\boldsymbol{1}\,; & / \end{matrix} \right) \#\, {}_pF_q(\hat{\mathcal{A}}_+ \hat{\mathcal{A}}_-) \# |z><z| = \Gamma\left(\tilde{\boldsymbol{b}} / \tilde{\boldsymbol{a}}\right) \#\, {}_pF_q(\hat{\mathcal{A}}_+ \hat{\mathcal{A}}_-) \# \tag{3.27}$$

The proof is simple. Using the expansion of CSs according to the Fock vectors, because the angular integral is $\left(|z|^2\right)^l \delta_{l\,s}$, we can verify this relation as follows

$$\sum_{l=0}^{\infty} \frac{\left(\hat{\mathcal{A}}_+\right)^l}{\rho(l)} \sum_{s=0}^{\infty} \frac{\left(\hat{\mathcal{A}}_-\right)^s}{\rho(s)} |0><0| \int \frac{d^2 z}{\pi} G^{q+1,0}_{p,\,q+1}\left( |z|^2 \middle| \begin{matrix} / \,; & \tilde{\boldsymbol{a}}\,\text{-}\,\boldsymbol{1} \\ 0\,, \tilde{\boldsymbol{b}}\,\text{-}\,\boldsymbol{1}\,; & / \end{matrix} \right) z^l \left(z^*\right)^s = \Gamma\left(\tilde{\boldsymbol{b}} / \tilde{\boldsymbol{a}}\right) \tag{3.28}$$

$$\begin{aligned} &\# \sum_{l=0}^{\infty} \frac{\left(\hat{\mathcal{A}}_+\right)^l}{\rho(l)} \sum_{s=0}^{\infty} \frac{\left(\hat{\mathcal{A}}_-\right)^s}{\rho(s)} \# \int_0^{\infty} d\left(|z|^2\right) G^{q+1,0}_{p,\,q+1}\left( |z|^2 \middle| \begin{matrix} / \,; & \tilde{\boldsymbol{a}}\,\text{-}\,\boldsymbol{1} \\ 0\,, \tilde{\boldsymbol{b}}\,\text{-}\,\boldsymbol{1}\,; & / \end{matrix} \right) \int_0^{2\pi} \frac{d\varphi}{2\pi} z^l \left(z^*\right)^s = \\ &= \Gamma\left(\tilde{\boldsymbol{b}} / \tilde{\boldsymbol{a}}\right) \#\, {}_pF_q(\hat{\mathcal{A}}_+ \hat{\mathcal{A}}_-) \# \end{aligned} \tag{3.29}$$

Let us also focus our attention on an integral that results from the decomposition relations of the unit operator and which will be useful to us further:

$$\int \frac{d^2 z}{\pi} G_{p,\,q+1}^{q+1,0}\left( |\,z\,|^2 \left| \begin{array}{cc} / \;; & \tilde{\boldsymbol{a}} \textbf{ - 1} \\ 0 \,, \tilde{\boldsymbol{b}} \textbf{ - 1} \;; & / \end{array} \right. \right) \left(|\,z\,|^2\right)^n = \Gamma\left(\tilde{\boldsymbol{b}} \,/ \tilde{\boldsymbol{a}}\right) \rho(n) \tag{3.30}$$

This results is a consequence of the fact that the Meijer's G-function satisfies the following classical integral [14]:

$$\int_0^\infty dx\, x^{\alpha-1}\, G_{p,q}^{m,n}\left( \beta x \left| \begin{array}{cc} \{a_i\}_1^n \;; & \{a_i\}_{n+1}^p \\ \{b_j\}_1^m \;; & \{b_j\}_{m+1}^q \end{array} \right. \right) = \frac{1}{\beta^\alpha} \frac{\prod_{j=1}^m \Gamma(b_j+\alpha) \prod_{i=1}^n \Gamma(1-a_i-\alpha)}{\prod_{j=m+1}^q \Gamma(1-b_j-\alpha) \prod_{i=n+1}^p \Gamma(a_i+\alpha)} \tag{3.31}$$

By analogy with the displacement operator defined by the canonical operators $\hat{\boldsymbol{a}}^+$ and $\hat{\boldsymbol{a}}$, the generalized exponential displacement operator, corresponding to the ladder operators $\hat{\mathcal{A}}_+$ and $\hat{\mathcal{A}}_-$, can also be defined as follows:

$$\hat{\mathcal{D}}(z) = \#\exp\left(z\hat{\mathcal{A}}_+ - z^*\hat{\mathcal{A}}_-\right)\# \tag{3.32}$$

At the end of this section, let us point out that, besides Barut-Girardello type CSs, other type of CSs can also be defined in the so called Klauder-Perelomov manner, as a result of the action of the generalized displacement operator on the vacuum state:

$$|\,z\,>_{KP} = \frac{1}{\sqrt{N_{KP}\left(|\,z\,|^2\right)}} \hat{\mathcal{D}}(z)\,|\,0\,> = \frac{1}{\sqrt{N_{KP}\left(|\,z\,|^2\right)}} \exp\left(z\hat{\mathcal{A}}_+\right)|\,0\,> \tag{3.33}$$

where we took into account that $\hat{\mathcal{A}}_-|\,0\,> = 0$.

Expanding the exponential in a power series and taking into account the result of the action of the operator $\hat{\mathcal{A}}_+$ on the vacuum state, we obtain that the set of CSs of Klauder-Perelomov type has the expression

$$|\,z\,>_{KP} = \frac{1}{\sqrt{{}_qF_p(\tilde{\boldsymbol{b}}\,;\tilde{\boldsymbol{a}};|\,z\,|^2)}} \sum_{n=0}^\infty \frac{z^n}{\sqrt{\rho_{KP}(n)}} |\,n\,> = \frac{1}{\sqrt{{}_qF_p(\tilde{\boldsymbol{b}}\,;\tilde{\boldsymbol{a}};|\,z\,|^2)}}\, {}_qF_p(\tilde{\boldsymbol{b}}\,;\tilde{\boldsymbol{a}};\,z\hat{\mathcal{A}}_+)|\,0\,> \tag{3.34}$$

The notation used here is

$$\rho_{KP}(n) = \frac{(n\,!)^2}{\rho(n)} \tag{3.35}$$

It can be seen that between the sets of CSs of the Barut-Giradello type $\{|\,z\,>\}$, on the one hand, and the Klauder-Perelomov $\{|\,z\,>_{KP}\}$, on the other hand, there is a dualism. This dualism consists mainly of the interchange of the indices $p$ and $q$, as well as of the sets of numbers $\{a_i\}$ and $\{b_j\}$ , that is, instead of the hypergeometric function ${}_pF_q(\tilde{\boldsymbol{a}};\tilde{\boldsymbol{b}};x)$, the function ${}_qF_p(\tilde{\boldsymbol{b}};\tilde{\boldsymbol{a}};x)$ appears, as well as instead of the structure function $\rho(n)$, the function $\rho_{KP}(n)$ apears [15].

## 4. Generalized displacement operator

Previously we have seen that the generalized CSs can also be obtained by applying the generalized hypergeometric function ${}_pF_q(z\hat{\mathcal{A}}_+)$ on the vacuum ket vector $|0>$, see, Eq. (3.17).

This idea appeared for the first time, for a particular case of pseudoharmonic oscillator, in Mojaveri and Dehghani's paper [16].

On the other hand, any function that has as its argument the annihilation operator, therefore also ${}_pF_q(z^*\hat{\mathcal{A}}_-)$ , applied to the vector $|0>$ leaves it unchanged, i.e.

$$ {}_pF_q(z^*\hat{\mathcal{A}}_-)|0>=|0> \tag{4.01}$$

So, it is obvious that a generalized displacement operator can be constructed, similar to the "usual" one $\hat{\mathcal{D}}(z)=e^{-\frac{1}{2}|z|^2}e^{z\hat{a}^+-z^*\hat{a}}$. Consequently, by analogy, we can define the *generalized analogue of the displacement operator* as follows:

$$\hat{\mathfrak{D}}(z)\equiv\frac{1}{\sqrt{{}_pF_q(|z|^2)}}\#{}_pF_q(z\hat{\mathcal{A}}_+)\;{}_pF_q(\pm z^*\hat{\mathcal{A}}_-)\# \tag{4.02}$$

As a consequence, the following relationship will have to be valid at the limit:

$$\lim_{\substack{p=q\\ \boldsymbol{a}=\boldsymbol{b}}}\hat{\mathfrak{D}}(z)=\hat{\mathcal{D}}(z) \tag{4.03}$$

In the calculations below we can drop the function ${}_pF_q(\pm z^*\hat{\mathcal{A}}_-)$, so we will consider that the generalized displacement operator has the expression (which agrees with [16]):

$$\hat{\mathfrak{D}}(z)\equiv\frac{1}{\sqrt{{}_pF_q(|z|^2)}}\,{}_pF_q(z\hat{\mathcal{A}}_+) \tag{4.04}$$

Using the definition of the CSs (and their complex conjugate counterpart), let's highlight another property of the generalized displacement operator, which can be considered as their eigenvalue equation

$$ {}_pF_q(z^*\hat{\mathcal{A}}_-)|\sigma>=\sum_{n=0}^{\infty}\frac{(z^*)^n}{\rho(n)}\left(\hat{\mathcal{A}}_-\right)^n|\sigma>=\sum_{n=0}^{\infty}\frac{(z^*)^n}{\rho(n)}\sigma^n|\sigma>={}_pF_q(z^*\sigma)|\sigma> $$
$$ <\sigma|{}_pF_q(z\hat{\mathcal{A}}_+)=<\sigma|\,{}_pF_q(z\sigma^*) \qquad (4.05) $$

It means that the function ${}_pF_q(z^*\sigma)$ is the eigenvalue of the operator ${}_pF_q(z^*\hat{\mathcal{A}}_-)$ attached to the eigenfunction $|\sigma>$ and similar for their complex conjugate.

The diagonal elements of the generalized displacement operator in CSs representation are

$$ <\sigma|\hat{\mathcal{D}}(z)|\sigma>=\frac{1}{\sqrt{{}_pF_q(|z|^2)}}\#<\sigma|_p F_q(z\hat{\mathcal{A}}_+)\,{}_pF_q(z^*\hat{\mathcal{A}}_-)|\sigma>\#= $$
$$ =\frac{1}{\sqrt{{}_pF_q(|z|^2)}}\frac{{}_pF_q(z\sigma^*)\,{}_pF_q(z^*\sigma)}{{}_pF_q(|\sigma|^2)} \qquad (4.06) $$

On the other hand, from the DOOT rules we have

$$ |l><l|=\frac{1}{\rho(l)}\#\left(\hat{\mathcal{A}}_+\right)^l|0><0|\left(\hat{\mathcal{A}}_-\right)^l\#=\frac{1}{\#{}_pF_q(\hat{\mathcal{A}}_+\hat{\mathcal{A}}_-)\#}\frac{\#\left(\hat{\mathcal{A}}_+\hat{\mathcal{A}}_-\right)^l\#}{\rho(l)} \qquad (4.07) $$

and it can be easy verified that this expression fulfill the completeness relation

$$ \sum_{l=0}^{\infty}|l><l|=1 \qquad (4.08) $$

Now, we will be able to write, which means that the displacement operator applied to the vacuum ket vector $|0>$ transforms it into a vector corresponding to the generalized CSs $|z>$.

$$ |z>=\hat{\mathcal{D}}(z)|0> \qquad (4.09) $$

For the moment, for reasons of clarity of exposition, we will return to the initial notation for generalized hypergeometric functions, ${}_pF_q(\tilde{\boldsymbol{a}}\,;\tilde{\boldsymbol{b}}\,;x)$.

Previously we saw that CSs satisfy the decomposition relation of the unit operator. Let us see how generalized states of displacement are involved in this relation, also taking into account the DOOT rules.

$$ \int d\mu(z)|z><z|=\int d\mu(z)\#\hat{\mathcal{D}}(z)|0><0|\hat{\mathcal{D}}^+(z)\#= $$
$$ =\frac{1}{\#{}_pF_q(\tilde{\boldsymbol{a}}\,;\tilde{\boldsymbol{b}}\,;\hat{\mathcal{A}}_+\hat{\mathcal{A}}_-)\#}\int d\mu(z)\#\hat{\mathcal{D}}(z)\hat{\mathcal{D}}^+(z)\#=\hat{I} \qquad (4.10) $$

From this follows a new property of the generalized displacement operator:

$$ \int d\mu(z)\#\hat{\mathcal{D}}(z)\hat{\mathcal{D}}^+(z)\#=\#{}_pF_q(\tilde{\boldsymbol{a}}\,;\tilde{\boldsymbol{b}}\,;\hat{\mathcal{A}}_+\hat{\mathcal{A}}_-)\# \qquad (4.11) $$

or, more explicitly, using the expression of integration measure

$$\int \frac{d^2 z}{\pi} G_{p,\,q+1}^{q+1,0}\left( |\,z\,|^2 \left| \begin{matrix} /\,; & \tilde{\boldsymbol{a}}\,\boldsymbol{-1} \\ 0\,,\tilde{\boldsymbol{b}}\,\boldsymbol{-1}\,; & / \end{matrix} \right. \right) {}_pF_q(\tilde{\boldsymbol{a}}\,;\tilde{\boldsymbol{b}}\,;|\,z\,|^2)\#\hat{\mathcal{D}}(z)\hat{\mathcal{D}}^{+}(z)\# = \\ = \Gamma\left(\tilde{\boldsymbol{b}}\,/\,\tilde{\boldsymbol{a}}\right)\#\,{}_pF_q(\tilde{\boldsymbol{a}}\,;\tilde{\boldsymbol{b}}\,;\hat{\mathcal{A}}_{+}\hat{\mathcal{A}}_{-})\# \tag{4.12}$$

Moreover, the CSs can also be defined as follows:

$$|z>=\hat{\mathcal{D}}(z)|\,0>=\frac{1}{\sqrt{{}_pF_q(\tilde{\boldsymbol{a}}\,;\tilde{\boldsymbol{b}}\,;|\,z\,|^2)}}\,{}_pF_q(\tilde{\boldsymbol{a}}\,;\tilde{\boldsymbol{b}}\,;z\hat{\mathcal{A}}_{+})|\,0> \tag{4.13}$$

$$<z\,|=<0\,|\,\hat{\mathcal{D}}^{+}(z)=\frac{1}{\sqrt{{}_pF_q(\tilde{\boldsymbol{a}}\,;\tilde{\boldsymbol{b}}\,;|\,z\,|^2)}}<0|\,{}_pF_q(\tilde{\boldsymbol{a}}\,;\tilde{\boldsymbol{b}}\,;z^{*}\hat{\mathcal{A}}_{-}) \tag{4.14}$$

Substituting into the decomposition relation the unit operator, and taking into account the DOOT rules, we obtain

$$\int d\mu(z)|\,z><z\,|= \\ =\frac{1}{\#\,{}_pF_q(\tilde{\boldsymbol{a}}\,;\tilde{\boldsymbol{b}}\,;\hat{\mathcal{A}}_{+}\hat{\mathcal{A}}_{-})\#}\int d\mu(z)\frac{1}{\sqrt{{}_pF_q(|\,z\,|^2)}}\#\,{}_pF_q(\tilde{\boldsymbol{a}}\,;\tilde{\boldsymbol{b}}\,;z\hat{\mathcal{A}}_{+})\,{}_pF_q(\tilde{\boldsymbol{a}}\,;\tilde{\boldsymbol{b}}\,;z^{*}\hat{\mathcal{A}}_{-})\# = \hat{I} \tag{4.15}$$

After substituting the expression of the integration measure, we will obtain an important relationship:

$$\int \frac{d^2 z}{\pi} G_{p,\,q+1}^{q+1,0}\left( |\,z\,|^2 \left| \begin{matrix} /\,; & \tilde{\boldsymbol{a}}\,\boldsymbol{-1} \\ 0\,,\tilde{\boldsymbol{b}}\,\boldsymbol{-1}\,; & / \end{matrix} \right. \right)\#\,{}_pF_q(\tilde{\boldsymbol{a}}\,;\tilde{\boldsymbol{b}}\,;z\hat{\mathcal{A}}_{+})\,{}_pF_q(\tilde{\boldsymbol{a}}\,;\tilde{\boldsymbol{b}}\,;z^{*}\hat{\mathcal{A}}_{-})\# = \\ = \Gamma\left(\tilde{\boldsymbol{b}}\,/\tilde{\boldsymbol{a}}\right)\#\,{}_pF_q(\tilde{\boldsymbol{a}}\,;\tilde{\boldsymbol{b}}\,;\hat{\mathcal{A}}_{+}\hat{\mathcal{A}}_{-})\# \tag{4.16}$$

With the help of this integral, another integral can be calculated in which the product $z^n\left(z^*\right)^m$ occurs.

$$\mathrm{Int}_{z z^*} \equiv \int \frac{d^2 z}{\pi} G_{p,\,q+1}^{q+1,0}\left( |\,z\,|^2 \left| \begin{matrix} /\,; & \tilde{\boldsymbol{a}}\,\boldsymbol{-1} \\ 0\,,\tilde{\boldsymbol{b}}\,\boldsymbol{-1}\,; & / \end{matrix} \right. \right) z^n\left(z^*\right)^m \#\,{}_pF_q(\tilde{\boldsymbol{a}}\,;\tilde{\boldsymbol{b}}\,;z\hat{\mathcal{A}}_{+})\,{}_pF_q(\tilde{\boldsymbol{a}}\,;\tilde{\boldsymbol{b}}\,;z^{*}\hat{\mathcal{A}}_{-}\# \tag{4.17}$$

Taking into account the equalities

$$\frac{\partial}{\partial\hat{\mathcal{A}}_{+}} = z\frac{\partial}{\partial\left(z\hat{\mathcal{A}}_{+}\right)}\ , \quad \frac{\partial}{\partial\hat{\mathcal{A}}_{-}} = z^{*}\frac{\partial}{\partial\left(z^{*}\hat{\mathcal{A}}_{-}\right)} \tag{4.18}$$

as well as the differentiation rule for hypergeometric functions [14]

$$\left(\frac{\partial}{\partial\,\hat{\mathcal{A}}_{+}}\right)^n {}_pF_q(\tilde{\boldsymbol{a}}\,;\tilde{\boldsymbol{b}}\,;z\hat{\mathcal{A}}_{+}) = z^n\frac{\prod_{i=1}^{p}\left(a_i\right)_n}{\prod_{j=1}^{q}\left(b_j\right)_n}\,{}_pF_q(\tilde{\boldsymbol{a}}+\boldsymbol{n}\,;\tilde{\boldsymbol{b}}+\boldsymbol{n};z\hat{\mathcal{A}}_{+}) \tag{4.19}$$

Consequently, we have

$$z^n \,{}_pF_q(\tilde{\boldsymbol{a}}\,;\tilde{\boldsymbol{b}}\,;z\,A_+)=\frac{\prod_{j=1}^{q}\left(1-b_j\right)_n}{\prod_{i=1}^{p}\left(1-a_i\right)_n}\left(\frac{\partial}{\partial\hat{\boldsymbol{\mathcal{A}}}_+}\right)^n {}_pF_q(\tilde{\boldsymbol{a}}-\boldsymbol{n}\,;\tilde{\boldsymbol{b}}-\boldsymbol{n};z\hat{\boldsymbol{\mathcal{A}}}_+) \qquad (4.20)$$

as well as their complex conjugate. Here we used the equality [17]

$$\left(a-n\right)_n=\left(-1\right)^n\left(1-a\right)_n \qquad (4.21)$$

The integral then becomes

$$\mathrm{Int}_{zz^*}=\frac{\prod_{j=1}^{q}\left(1-b_j\right)_n}{\prod_{i=1}^{p}\left(1-a_i\right)_n}\frac{\prod_{j=1}^{q}\left(1-b_j\right)_m}{\prod_{i=1}^{p}\left(1-a_i\right)_m}\left(\frac{\partial}{\partial\hat{\boldsymbol{\mathcal{A}}}_+}\right)^n\left(\frac{\partial}{\partial\hat{\boldsymbol{\mathcal{A}}}_-}\right)^m\times$$
$$\times\int\frac{d^2z}{\pi}G_{p,\,q+1}^{q+1,0}\left(|\,z\,|^2\middle|\begin{matrix}/\,; & \tilde{\boldsymbol{a}}\,\boldsymbol{-1}\\ 0\,,\tilde{\boldsymbol{b}}\,\boldsymbol{-1}\,; & /\end{matrix}\right){}_pF_q(\tilde{\boldsymbol{a}}-\boldsymbol{n}\,;\tilde{\boldsymbol{b}}-\boldsymbol{n};z\hat{\boldsymbol{\mathcal{A}}}_+)\,{}_pF_q(\tilde{\boldsymbol{a}}-\boldsymbol{n}\,;\tilde{\boldsymbol{b}}-\boldsymbol{n};z^*\hat{\boldsymbol{\mathcal{A}}}_-) \qquad (4.22)$$

If we write the hypergeometric functions as the series and take into account that the angular integral has the value $\left(|\,z\,|^2\right)^l\delta_{l\,s}$ , as well as that the integral over the variable $|\,z\,|^2$ is

$$\int_0^\infty d\left(|\,z\,|^2\right)G_{p,\,q+1}^{q+1,0}\left(|\,z\,|^2\middle|\begin{matrix}/\,; & \tilde{\boldsymbol{a}}\,\boldsymbol{-1}\\ 0\,,\tilde{\boldsymbol{b}}\,\boldsymbol{-1}\,; & /\end{matrix}\right)\left(|\,z\,|^2\right)^l=\Gamma\left(\tilde{\boldsymbol{b}}\,/\tilde{\boldsymbol{a}}\right)\rho\left(l\right) \qquad (4.23)$$

then the integral finally becomes

$$\mathrm{Int}_{zz^*}=\Gamma\left(\tilde{\boldsymbol{b}}\,/\tilde{\boldsymbol{a}}\right)\frac{\prod_{j=1}^{q}\left(1-b_j\right)_n}{\prod_{i=1}^{p}\left(1-a_i\right)_n}\frac{\prod_{j=1}^{q}\left(1-b_j\right)_m}{\prod_{i=1}^{p}\left(1-a_i\right)_m}\times$$
$$\times\left(\frac{\partial}{\partial\hat{\boldsymbol{\mathcal{A}}}_+}\right)^n\left(\frac{\partial}{\partial\hat{\boldsymbol{\mathcal{A}}}_-}\right)^m\sum_{l=0}^{\infty}\frac{\prod_{i=1}^{p}\left(a_i-n\right)_l}{\prod_{j=1}^{q}\left(b_j-n\right)_l}\frac{\prod_{i=1}^{p}\left(a_i-m\right)_l}{\prod_{j=1}^{q}\left(b_j-m\right)_l}\frac{\prod_{j=1}^{q}\left(b_j\right)_l}{\prod_{i=1}^{p}\left(a_i\right)_l}\frac{\left(\hat{\boldsymbol{\mathcal{A}}}_+\hat{\boldsymbol{\mathcal{A}}}_-\right)^l}{l!} \qquad (4.24)$$

The above sum is a hypergeometric function, so the final result can also be written as:

$$\int\frac{d^2z}{\pi}G_{p,\,q+1}^{q+1,0}\left(|\,z\,|^2\middle|\begin{matrix}/\,; & \tilde{\boldsymbol{a}}\,\boldsymbol{-1}\\ 0\,,\tilde{\boldsymbol{b}}\,\boldsymbol{-1}\,; & /\end{matrix}\right)z^n\left(z^*\right)^m\#\,{}_pF_q(\tilde{\boldsymbol{a}}\,;\tilde{\boldsymbol{b}}\,;z\hat{\boldsymbol{\mathcal{A}}}_+)\,{}_pF_q(\tilde{\boldsymbol{a}}\,;\tilde{\boldsymbol{b}}\,;z^*\hat{\boldsymbol{\mathcal{A}}}_-)\#=$$
$$=\Gamma\left(\tilde{\boldsymbol{b}}\,/\tilde{\boldsymbol{a}}\right)\frac{\prod_{j=1}^{q}\left(1-b_j\right)_n}{\prod_{i=1}^{p}\left(1-a_i\right)_n}\frac{\prod_{j=1}^{q}\left(1-b_j\right)_m}{\prod_{i=1}^{p}\left(1-a_i\right)_m}\left(\frac{\partial}{\partial\hat{\boldsymbol{\mathcal{A}}}_+}\right)^n\left(\frac{\partial}{\partial\hat{\boldsymbol{\mathcal{A}}}_-}\right)^m\#\,{}_pF_q(\tilde{\boldsymbol{a}}\,\boldsymbol{-n},\tilde{\boldsymbol{a}}\,\boldsymbol{-n}\,,\tilde{\boldsymbol{b}}\,;\tilde{\boldsymbol{b}}\,\boldsymbol{-n},\tilde{\boldsymbol{b}}\,\boldsymbol{-n}\,,\tilde{\boldsymbol{a}}\,;\hat{\boldsymbol{\mathcal{A}}}_+\hat{\boldsymbol{\mathcal{A}}}_-)\#$$

(4.25)

For the particular case $p = q = 0$ and $\tilde{\boldsymbol{a}} = \tilde{\boldsymbol{b}}$ , in which case and $\hat{\mathcal{A}}_{-} = \hat{\boldsymbol{a}}$, and $\hat{\mathcal{A}}_{+} = \hat{\boldsymbol{a}}^{+}$ we obtain

$$\int \frac{d^2 z}{\pi} e^{-|z|^2} z^n \left(z^*\right)^m e^{z\boldsymbol{a}^+} e^{z^*\hat{\boldsymbol{a}}} = \left(\frac{\partial}{\partial \hat{\boldsymbol{a}}^+}\right)^n \left(\frac{\partial}{\partial \hat{\boldsymbol{a}}}\right)^m e^{\hat{\boldsymbol{a}}^+ \hat{\boldsymbol{a}}} \tag{4.26}$$

The result is verified if $n = m = 0$, thus we will arrive at the mathematical relation below.

For the particular case of canonical CSs, for which $\tilde{\boldsymbol{b}} = \tilde{\boldsymbol{a}}$ and $p = q$, the above relation reduces to a well-known mathematical relationship [18]:

$$\int \frac{d^2 z}{\pi} e^{-|z|^2 + z\hat{\boldsymbol{a}}^+ + z^*\hat{\boldsymbol{a}}} = e^{\hat{\boldsymbol{a}}^+ \hat{\boldsymbol{a}}} \tag{4.27}$$

or in an equivalent form

$$\int \frac{d^2 z}{\pi} e^{-(\hat{\boldsymbol{a}}^+ - z)(\hat{\boldsymbol{a}} - z^*)} = 1 \tag{4.28}$$

Replacing the expression of the integration measure, we will have, successively:

$$\frac{1}{\#_p F_q(\tilde{\boldsymbol{a}}\,;\tilde{\boldsymbol{b}}\,;\hat{\mathcal{A}}_+\hat{\mathcal{A}}_-)\#} \frac{\prod_{i=1}^{p}\Gamma(a_i)}{\prod_{j=1}^{q}\Gamma(b_j)} \frac{d\varphi}{2\pi} \int_0^\infty d(|z|^2)\, G_{p,\,q+1}^{q+1,0}\left(|z|^2 \middle| \begin{array}{cc} /\,; & \tilde{\boldsymbol{a}}\,\boldsymbol{-1} \\ 0\,,\tilde{\boldsymbol{b}}\,\boldsymbol{-1}\,; & / \end{array}\right) \times$$
$$\times \int_0^{2\pi} \frac{d\varphi}{2\pi} \#_p F_q(\tilde{\boldsymbol{a}}\,;\tilde{\boldsymbol{b}}\,;z\hat{\mathcal{A}}_+)\,_p F_q(\tilde{\boldsymbol{a}}\,;\tilde{\boldsymbol{b}}\,;z^*\hat{\mathcal{A}}_-)\# = \hat{I} \tag{4.29}$$

The angular integral has the following expression:

$$\int_0^{2\pi} \frac{d\varphi}{2\pi} \#_p F_q(\tilde{\boldsymbol{a}}\,;\tilde{\boldsymbol{b}}\,;z\hat{\mathcal{A}}_+)\,_p F_q(\tilde{\boldsymbol{a}}\,;\tilde{\boldsymbol{b}}\,;z^*\hat{\mathcal{A}}_-)\# = \sum_{n=0}^{\infty} \frac{\#\left(\hat{\mathcal{A}}_+\hat{\mathcal{A}}_-\right)^n\#}{\left[\rho(n)\right]^2} \left(|z|^2\right)^n \tag{4.30}$$

and the result of the integral after $|z|^2$ is

$$\int_0^\infty d(|z|^2)\, G_{p,\,q+1}^{q+1,0}\left(|z|^2 \middle| \begin{array}{cc} /\,; & \tilde{\boldsymbol{a}}\,\boldsymbol{-1} \\ 0\,,\tilde{\boldsymbol{b}}\,\boldsymbol{-1}\,; & / \end{array}\right)\left(|z|^2\right)^n = \Gamma\left(\tilde{\boldsymbol{b}}\,/\tilde{\boldsymbol{a}}\right)\rho(n) \tag{4.31}$$

which means that equality is satisfied.

Consequently, the generalized displacement operators satisfy the relation

$$\int d\mu(z) \#\hat{\mathcal{D}}(z)\,\hat{\mathcal{D}}^+(z)\# = \#_p F_q(\tilde{\boldsymbol{a}}\,;\tilde{\boldsymbol{b}}\,;\hat{\mathcal{A}}_+\hat{\mathcal{A}}_-)\# \tag{4.32}$$

Considering the DOOT rules, according to which the creation and annihilation operators commute, so that in the end we have an ordered product of the operators, we can easily verify the validity of the properties below.

The generalized displacement operators are unitary operators. This can be shown by considering the scalar product of self-consistent states:

$$<z|z>=<0|\#\hat{\mathcal{D}}^{+}(z)\hat{\mathcal{D}}(z)|\#|0>=1 \tag{4.33}$$

To achieve equality, the condition must be met

$$\#\hat{\mathcal{D}}^{+}(z)\hat{\mathcal{D}}(z)|\#=\#\hat{\mathcal{D}}(z)\hat{\mathcal{D}}^{+}(z)|\#=1 \tag{4.34}$$

On the other hand, we can write

$$<z|z>=\frac{1}{{}_pF_q(\tilde{\boldsymbol{a}};\tilde{\boldsymbol{b}};|z|^2)}<0|\#_pF_q(\tilde{\boldsymbol{a}};\tilde{\boldsymbol{b}};z^*\hat{\mathcal{A}}_-)\,{}_pF_q(\tilde{\boldsymbol{a}};\tilde{\boldsymbol{b}};z\hat{\mathcal{A}}_+)\#|0>=$$

$$=\frac{1}{{}_pF_q(\tilde{\boldsymbol{a}};\tilde{\boldsymbol{b}};|z|^2)}\sum_{n=0}^{\infty}\frac{(z^*)^n}{\rho(n)}\#<0|\left(\hat{\mathcal{A}}_-\right)^n\sum_{m=0}^{\infty}\frac{z^m}{\rho(m)}\left(\hat{\mathcal{A}}_+\right)^m|0>\#= \tag{4.35}$$

$$=\frac{1}{{}_pF_q(\tilde{\boldsymbol{a}};\tilde{\boldsymbol{b}};|z|^2)}\sum_{n=0}^{\infty}\frac{(z^*)^n}{\sqrt{\rho(n)}}\sum_{m=0}^{\infty}\frac{z^m}{\sqrt{\rho(m)}}<n|m>=\frac{1}{{}_pF_q(\tilde{\boldsymbol{a}};\tilde{\boldsymbol{b}};|z|^2)}\sum_{n=0}^{\infty}\frac{(|z|^2)^n}{\rho(n)}=1$$

Also using the DOOT rules, namely the commutative property of ladder operators, we can write

$$\begin{aligned}&\#\hat{\mathcal{D}}^{+}(z)\hat{\mathcal{A}}_+\hat{\mathcal{D}}(z)\#=\hat{\mathcal{A}}_+\#\hat{\mathcal{D}}^{+}(z)\hat{\mathcal{D}}(z)\#=\hat{\mathcal{A}}_+,\\&\#\hat{\mathcal{D}}^{+}(z)\hat{\mathcal{A}}_-\hat{\mathcal{D}}(z)\#=\#\hat{\mathcal{D}}^{+}(z)\hat{\mathcal{D}}(z)\#\hat{\mathcal{A}}_-=\hat{\mathcal{A}}_-\end{aligned} \tag{4.36}$$

## 5. Construction of the displaced coherent states

Let's see what is the result of the successive application of two displacement operators, which depend on different arguments, respectively $\hat{\mathcal{D}}(\varepsilon\sigma)$ and $\hat{\mathcal{D}}(\lambda z)$ on the vacuum ket vector $|0>$. The goal is to obtain the expression of CSs that have as argument a linear combination of complex variables $z=|z|\exp(\mathrm{i}\varphi_z)$ and $\sigma=|\sigma|\exp(\mathrm{i}\varphi_\sigma)$, that is $Z=\varepsilon z+\lambda\sigma$, where $\varepsilon$ and $\lambda$ are real numbers.

This linear combination of complex numbers is, of course, also a complex number $Z$, whose modulus and phase are

$$\begin{aligned}&Z=\varepsilon z+\lambda\sigma=\mathcal{Re}Z+\mathrm{i}\,\mathcal{Im}Z=|Z|\exp(\mathrm{i}\varphi_Z)\\&|Z|^2=\varepsilon^2|z|^2+\lambda^2|\sigma|^2+2\varepsilon\lambda|z|\cdot|\sigma|\cos(\varphi_z-\varphi_\sigma)\\&\varphi_Z=\arctan\left(\frac{\varepsilon|z|\sin\varphi_z+\lambda|z|\sin\varphi_\sigma}{\varepsilon|z|\cos\varphi_z+\lambda|z|\cos\varphi_\sigma}\right)\end{aligned} \tag{5.01}$$

The expressions obtained will be, among others, useful for calculating the Wigner operator in the representation of CSs [18].

Consequently, CSs having as argument the linear combination of complex numbers is defined as follows:

$$| Z > \equiv | \varepsilon z + \lambda \sigma > = \hat{\mathfrak{D}}(\lambda \sigma)\, \hat{\mathfrak{D}}(\varepsilon z) | 0 > \qquad (5.02)$$

Let us see what is the expression of the development of these CSs in terms of the Fock vectors $| n >$. We will start from the definition relation of CSs with the argument $\varepsilon z$

$$\hat{\mathfrak{D}}(\varepsilon z) | 0 > = | \varepsilon z > \qquad (5.03)$$

and we will act on it with the displacement operator $\hat{\mathfrak{D}}(\lambda \sigma)$. We obtain, successively

$$\begin{aligned} \hat{\mathfrak{D}}(\lambda \sigma) | \varepsilon z > &= \frac{1}{\sqrt{{}_pF_q(|\lambda \sigma|^2)}} \frac{1}{\sqrt{{}_pF_q(\varepsilon |z|^2)}} \# {}_pF_q(\lambda \sigma \hat{\mathcal{A}}_+)\, {}_pF_q(\varepsilon z \hat{\mathcal{A}}_+) \# | 0 > = \\ &= \frac{1}{\sqrt{{}_pF_q(|\lambda \sigma|^2)}}\, {}_pF_q(\lambda \sigma \hat{\mathcal{A}}_+) | \varepsilon z > \end{aligned} \qquad (5.04)$$

$${}_pF_q(\lambda \sigma \hat{\mathcal{A}}_+) | \varepsilon z > = \frac{1}{\sqrt{{}_pF_q(|\varepsilon z|^2)}} \sum_{m=0}^{\infty} \frac{(\lambda \sigma)^m}{\rho(m)} \left[ \sum_{n=0}^{\infty} \frac{(\varepsilon z)^n}{\sqrt{\rho(n)}} \left(\hat{\mathcal{A}}_+\right)^m | n > \right] \qquad (5.05)$$

$$\left(\hat{\mathcal{A}}_+\right)^m | n > = \sqrt{\frac{\rho(n+m)}{\rho(n)}}\, | n+m > \qquad (5.06)$$

Substituting above and using a new summation index $l = n + m$, we obtain

$$\begin{aligned} {}_pF_q(\lambda\, \sigma \hat{\mathcal{A}}_+) | \varepsilon z > &= \frac{1}{\sqrt{{}_pF_q(|\varepsilon z|^2)}} \sum_{n=0}^{\infty} \frac{(\varepsilon z)^n}{\sqrt{\rho(n)}} \sum_{m=0}^{\infty} \frac{(\lambda \sigma)^m}{\rho(m)} \sqrt{\frac{\rho(n+m)}{\rho(n)}}\, | n+m > = \\ &= \frac{1}{\sqrt{{}_pF_q(|\varepsilon z|^2)}} \sum_{l=m}^{\infty} \frac{1}{\sqrt{\rho(l)}} \left[ \sum_{m=0}^{l} \frac{\rho(l)}{\rho(m)\rho(l-m)} (\varepsilon z)^{l-m} (\lambda \sigma)^m \right] | l > = \\ &= \frac{1}{\sqrt{{}_pF_q(|\varepsilon z|^2)}} \sum_{l=0}^{\infty} \frac{1}{\sqrt{\rho(l)}} \left[ \sum_{m=0}^{l} \begin{pmatrix} \rho(l) \\ \rho(m) \end{pmatrix} (\varepsilon z)^{l-m} (\lambda \sigma)^m \right] | l > \end{aligned} \qquad (5.07)$$

In addition, since $m \le l$ and the sum over the index $l$ no longer depends on the index $m$, we can take $m = 0$.

The quantity between square brackets represents precisely the *generalized expression of Newton's binomial*. For this quantity we will use square brackets:

$$[Z]^l \equiv [\varepsilon z + \lambda \sigma]^l = \sum_{m=0}^{l} \begin{pmatrix} \rho(l) \\ \rho(m) \end{pmatrix} (\varepsilon z)^{l-m} (\lambda \sigma)^m = \sum_{m=0}^{l} \begin{pmatrix} l \\ m \end{pmatrix} \begin{pmatrix} [g(l)!] \\ [g(m)!] \end{pmatrix} (\varepsilon z)^{l-m} (\lambda \sigma)^m \qquad (5.08)$$

So, a sum between square brackets raised to a power represents the generalized expression of the Newton's binomial:

$$\left[x+y\right]^{l} \equiv \sum_{m=0}^{l}\begin{pmatrix}\rho(l)\\ \rho(m)\end{pmatrix} x^{l-m} y^{m} = \sum_{m=0}^{l}\begin{pmatrix} l\\ m\end{pmatrix}\begin{pmatrix}[g(l)!]\\ [g(m)!]\end{pmatrix} x^{l-m} y^{m} \qquad (5.09)$$

Finally, after substitutions, we will get

$$\begin{aligned}\hat{\mathfrak{D}}(\lambda\,\sigma)|\,\varepsilon\, z>&=\frac{1}{\sqrt{{}_pF_q(|\,\lambda\,\sigma\,|^2)}}\frac{1}{\sqrt{{}_pF_q(|\,\varepsilon\, z\,|^2)}}\sum_{l=0}^{\infty}\frac{[Z]^{l}}{\sqrt{\rho(l)}}|\,l>=\\ &=\frac{\sqrt{{}_pF_q\left(|[Z]|^2\right)}}{\sqrt{{}_pF_q(|\lambda\sigma\,|^2)\,{}_pF_q(|\varepsilon\, z\,|^2)}}|\,Z>\end{aligned} \qquad (5.10)$$

where the generalized hypergeometric function depending on the displaced variable is

$${}_pF_q(\boldsymbol{a}\,;\boldsymbol{b};|\left[\varepsilon\, z+\lambda\sigma\right]|^2)\equiv {}_pF_q\left(\boldsymbol{a}\,;\boldsymbol{b};|[Z]|^2\right)$$

$$\begin{aligned}{}_pF_q\left(|[Z]|^2\right)&=\sum_{n=0}^{\infty}\frac{1}{\rho(n)}\left(|\left[\varepsilon\, z+\lambda\sigma\right]|^2\right)^n=\sum_{n=0}^{\infty}\frac{\prod_{i=1}^{p}(a_i)_n}{\prod_{j=1}^{q}(b_j)_n}\frac{\left(|\left[\varepsilon\, z+\lambda\sigma\right]|^2\right)^n}{n!}=\\ &=\sum_{n=0}^{\infty}\frac{\prod_{i=1}^{p}(a_i)_n}{\prod_{j=1}^{q}(b_j)_n}\frac{1}{n!}\sum_{m=0}^{n}\begin{pmatrix}\rho(n)\\ \rho(m)\end{pmatrix}(\varepsilon\, z)^{n-m}(\lambda\,\sigma)^m\end{aligned} \qquad (5.11)$$

The final expression for the shifted CSs, defined by the generalized displacement operators is then

$$|\,\varepsilon\, z+\lambda\sigma>=\sqrt{\frac{{}_pF_q(|\lambda\sigma\,|^2)\,{}_pF_q(|\varepsilon\, z\,|^2)}{{}_pF_q\left(|\left[\varepsilon\, z+\lambda\sigma\right]|^2\right)}}\hat{\mathfrak{D}}(\lambda\,\sigma)\hat{\mathfrak{D}}(\varepsilon\, z)|\,0> \qquad (5.12)$$

or, in equivalent manner, as their expansion with respect to the Fock vectors

$$|\,\varepsilon\, z+\lambda\sigma>=\frac{1}{\sqrt{{}_pF_q\left(|\left[\varepsilon\, z+\lambda\sigma\right]|^2\right)}}\sum_{n=0}^{\infty}\frac{\left[\varepsilon\, z+\lambda\sigma\right]^n}{\sqrt{\rho(n)}}|\,n> \qquad (5.13)$$

Finally, this is the expression for *generalized CSs with displaced (shifted) argument.*

This can also be written in the following form:

$$|\,\varepsilon\, z+\lambda\sigma>=\frac{1}{\sqrt{{}_pF_q\left(|\left[\varepsilon\, z+\lambda\sigma\right]|^2\right)}}\,{}_pF_q\left(\left[\varepsilon\, z+\lambda\sigma\right]\hat{\mathcal{A}}_+\right)|\,0> \qquad (5.14)$$

and similar for their complex conjugate counterpart.

Let's find the expression of the integration measure $d\mu(\varepsilon z+\lambda\sigma)=d\mu(Z)$ from the decomposition relation of the unit operator. For simplification, we will use the variable $Z=\varepsilon z+\lambda\sigma$, so that the CSs will be

$$|Z>=\frac{1}{\sqrt{{}_pF_q\left(|[Z]|^2\right)}}\sum_{n=0}^{\infty}\frac{[Z]^n}{\sqrt{\rho(n)}}|n>=\frac{1}{\sqrt{{}_pF_q\left(|[Z]|^2\right)}}\,{}_pF_q\left([Z]\hat{\mathcal{A}}_+\right)|0> \tag{5.15}$$

$$<Z|=\frac{1}{\sqrt{{}_pF_q\left(|[Z]|^2\right)}}\sum_{n=0}^{\infty}\frac{[Z^*]^n}{\sqrt{\rho(n)}}<n|=\frac{1}{\sqrt{{}_pF_q\left(|[Z]|^2\right)}}<0|\,{}_pF_q\left([Z^*]\hat{\mathcal{A}}_+\right) \tag{5.16}$$

and the decomposition relation of the unit operator will be written as

$$\int d\mu(Z)|Z><Z|=1 \tag{5.17}$$

Let's look for the expression of the integration measure in the following form

$$d\mu(Z)=\frac{d^2Z}{\pi}h\left(|[Z]|^2\right){}_pF_q\left(|[Z]|^2\right)=\frac{d\varphi_Z}{2\pi}d\left(|Z|^2\right)h\left(|[Z]|^2\right){}_pF_q\left(|[Z]|^2\right) \tag{5.18}$$

and we need to find the weighting function $h\left(|[Z]|^2\right)$.

$$\begin{gathered}\int\frac{d^2Z}{\pi}h\left(|[Z]|^2\right){}_pF_q\left(|[Z]|^2\right)\sum_{l=0}^{\infty}\frac{[Z]^l}{\sqrt{\rho(l)}}|l>\sum_{s=0}^{\infty}\frac{[Z^*]^s}{\sqrt{\rho(s)}}<s|=\\=\sum_{l=0}^{\infty}\sum_{s=0}^{\infty}\frac{|l><s|}{\sqrt{\rho(l)}\sqrt{\rho(s)}}\int_0^{\infty}d\left(|[Z]|^2\right)h\left(|[Z]|^2\right){}_pF_q\left(|[Z]|^2\right)\int_0^{2\pi}\frac{d\varphi_Z}{2\pi}[Z]^l[Z^*]^s=1\end{gathered} \tag{5.19}$$

The angular integral is non-zero only if we have equal exponents: $s=l$, which is in accordance with the condition of satisfying the closure relation of the Fock vectors $\sum_{l=0}^{\infty}|l><l|=1$. Therefore, we have:

$$\sum_{l=0}^{\infty}\frac{|l><l|}{\rho(l)}\int_0^{\infty}d\left(|[Z]|^2\right)h\left(|[Z]|^2\right){}_pF_q\left(|[Z]|^2\right)\left(|[Z]|^2\right)^l=1 \tag{5.20}$$

Reminding us of the definition of $\rho(l)$

$$\rho(l)=l!\frac{\prod_{j=1}^{q}(b_j)_l}{\prod_{i=1}^{p}(a_i)_l}=\Gamma\left(\tilde{\boldsymbol{a}}/\tilde{\boldsymbol{b}}\right)\Gamma(l+1)\frac{\prod_{j=1}^{q}\Gamma(b_j+l)}{\prod_{i=1}^{p}\Gamma(a_i+l)} \tag{5.21}$$

it is obvious that the above integral must have the value

$$\int_0^\infty d\left(|Z|^2\right) h\left(|[Z]|^2\right) {}_pF_q\left(|[Z]|^2\right) \left(|[Z]|^2\right)^l = \rho(l) = \Gamma\left(\tilde{\boldsymbol{a}} / \tilde{\boldsymbol{b}}\right) \Gamma(l+1) \frac{\prod_{j=1}^{q} \Gamma\left(b_j + l\right)}{\prod_{i=1}^{p} \Gamma\left(a_i + l\right)} \tag{5.22}$$

After changing the exponent $l = \alpha - 1$, the integral becomes

$$\int_0^\infty d\left(|[Z]|^2\right) h\left(|[Z]|^2\right) {}_pF_q\left(|[Z]|^2\right) \left(|[Z]|^2\right)^{\alpha-1} = \Gamma\left(\tilde{\boldsymbol{a}} / \tilde{\boldsymbol{b}}\right) \Gamma(\alpha) \frac{\prod_{j=1}^{q} \Gamma\left(b_j - 1 + \alpha\right)}{\prod_{i=1}^{p} \Gamma\left(a_i - 1 + \alpha\right)} \tag{5.23}$$

This is an integral of the type of Stieltjes moments problem and their solution imply the Meijer G-function:

$$h\left(|[Z]|^2\right) = \Gamma\left(\tilde{\boldsymbol{a}} / \tilde{\boldsymbol{b}}\right) G_{p,\,q+1}^{q+1,0}\left( |[Z]|^2 \left| \begin{matrix} / \;; & \{a_i - 1\}_1^p \\ 0\,, \{b_j - 1\}_1^q \;; & / \end{matrix} \right. \right) {}_pF_q\left(|[Z]|^2\right) \tag{5.24}$$

Consequently, the integration measure will have the final expression

$$d\mu(Z) = \Gamma\left(\tilde{\boldsymbol{a}} / \tilde{\boldsymbol{b}}\right) d\left(|[Z]|^2\right) \frac{d\varphi_Z}{2\pi} G_{p,\,q+1}^{q+1,0}\left( |[Z]|^2 \left| \begin{matrix} / \;; & \{a_i - 1\}_1^p \\ 0\,, \{b_j - 1\}_1^q \;; & / \end{matrix} \right. \right) {}_pF_q\left(|[Z]|^2\right) \tag{5.25}$$

In the notation of old variables, we will have

$$|\,\varepsilon z + \lambda\sigma >= \frac{1}{\sqrt{{}_pF_q\left(|\{\varepsilon z + \lambda\sigma\}|^2\right)}} \sum_{n=0}^{\infty} \frac{\left[\varepsilon z + \lambda\sigma\right]^n}{\sqrt{\rho(n)}} |\, n > \tag{5.26}$$

$$\begin{gathered} d\mu(\varepsilon z + \lambda\sigma) = \Gamma\left(\tilde{\boldsymbol{a}} / \tilde{\boldsymbol{b}}\right) d\left(|\,\varepsilon z + \lambda\sigma|^2\right) \frac{d\varphi_Z}{2\pi} \times \\ \times G_{p,\,q+1}^{q+1,0}\left( |\left[\varepsilon z + \lambda\sigma\right]|^2 \left| \begin{matrix} / \;; & \{a_i - 1\}_1^p \\ 0\,, \{b_j - 1\}_1^q \;; & / \end{matrix} \right. \right) {}_pF_q\left(|\left[\varepsilon z + \lambda\sigma\right]|^2\right) \end{gathered} \tag{5.27}$$

Let's evaluate the square of the modulus

$$\begin{gathered} \left(|[Z]|^2\right)^l = |\left[\varepsilon z + \lambda\sigma\right]|^2 = \left(\left[\varepsilon z + \lambda\sigma\right]\left[\varepsilon z^* + \lambda\sigma^*\right]\right)^l = \\ = \sum_{n=0}^{l} \begin{pmatrix} \rho(l) \\ \rho(n) \end{pmatrix} (\varepsilon z)^{l-n} (\lambda\sigma)^n \sum_{m=0}^{l} \begin{pmatrix} \rho(l) \\ \rho(m) \end{pmatrix} \left(\varepsilon z^*\right)^{l-m} \left(\lambda\sigma^*\right)^m = \\ = \sum_{n=0}^{l} \begin{pmatrix} \rho(l) \\ \rho(n) \end{pmatrix}^2 \left(\varepsilon^2 |z|^2\right)^{l-n} \left(\lambda^2 |\sigma|^2\right)^n \delta_{nm} \end{gathered} \tag{5.28}$$

The expression of the integration measure can be deduced much more simply if the DOOT rules are used.

If we substitute Eqs. (5.15) and (5.16) in Eq. (5.17), the decomposition relation of the unit operator will be

$$\int d\mu([Z]) \frac{1}{{}_pF_q\left(|[Z]|^2\right)} \#\, {}_pF_q([Z]\hat{\mathcal{A}}_+) |\, 0 >< 0 |\, {}_pF_q([Z^*]\hat{\mathcal{A}}_-) \# = $$
$$= \frac{1}{\#\, {}_pF_q(\hat{\mathcal{A}}_+\hat{\mathcal{A}}_-) \#} \int_0^\infty d\left(|Z|^2\right) h\left(|Z|^2\right) \int_0^{2\pi} \frac{d\varphi_Z}{2\pi} \#\, {}_pF_q([Z]\hat{\mathcal{A}}_+)\ {}_pF_q([Z^*]\hat{\mathcal{A}}_-) \# = 1 \tag{5.29}$$

The angular integral is

$$\int_0^{2\pi} \frac{d\varphi_Z}{2\pi} \#\, {}_pF_q([Z]\hat{\mathcal{A}}_+)\ {}_pF_q([Z^*]\hat{\mathcal{A}}_-) \# = \sum_{l=0}^\infty \# \frac{\left(\hat{\mathcal{A}}_+\right)^l}{\rho(l)} \sum_{s=0}^\infty \frac{\left(\hat{\mathcal{A}}_-\right)^s}{\rho(s)} \# \int_0^{2\pi} \frac{d\varphi_Z}{2\pi} [Z]^l [Z^*]^s = $$
$$= \sum_{l=0}^\infty \frac{\#\left(\hat{\mathcal{A}}_+\hat{\mathcal{A}}_-\right)^l \#}{[\rho(l)]^2} \left(|[Z]|^2\right)^l \delta_{l\,s} \tag{5.30}$$

The last angular integral has the value $\left(|[Z]|^2\right)^l \delta_{l\,s}$, so that the previous relationship becomes

$$\frac{1}{\#\, {}_pF_q(\hat{\mathcal{A}}_+\hat{\mathcal{A}}_-) \#} \sum_{l=0}^\infty \frac{\#\left(\hat{\mathcal{A}}_+\hat{\mathcal{A}}_-\right)^l \#}{[\rho(l)]^2} \int_0^\infty d\left(|[Z]|^2\right) h(|[Z]|) \left(|[Z]|^2\right)^l = 1 \tag{5.31}$$

and we obtain the same expression as Eq. (5. 22), and finally, the integration measure.

The decomposition relation of the unit operator will now be written as

$$\int d\mu([Z]) |Z >< Z| = $$
$$= \frac{1}{\#\, {}_pF_q(\hat{\mathcal{A}}_+\hat{\mathcal{A}}_-) \#} \int d\mu([Z]) \frac{1}{{}_pF_q\left(|[Z]|^2\right)} \#\, {}_pF_q([Z]\hat{\mathcal{A}}_+) |\, {}_pF_q([Z^*]\hat{\mathcal{A}}_-) \# = 1 \tag{5.32}$$

In the old variables, this becomes

$$\int \frac{d^2(\varepsilon z + \lambda\sigma)}{\pi} G_{p,\,q+1}^{q+1,0}\left( |\{\varepsilon z + \lambda\sigma\}|^2 \left| \begin{array}{cc} / \,; & \{a_i - 1\}_1^p \\ 0\,, \{b_j - 1\}_1^q \,; & / \end{array} \right. \right) \times$$
$$\times {}_pF_q\left([\lambda z + \varepsilon\sigma]\hat{\mathcal{A}}_+\right)\ {}_pF_q\left([\lambda z^* + \varepsilon\sigma^*]\hat{\mathcal{A}}_-\right) = \#\, {}_pF_q(\hat{\mathcal{A}}_+\hat{\mathcal{A}}_-) \# \tag{5.33}$$

In the context of what is indicated above, let us define the generalized displacement operator corresponding to the variable Z, as follows:

$$\hat{\mathcal{D}}(Z) = \frac{1}{\sqrt{{}_pF_q\left([Z]^2\right)}}\,{}_pF_q\left([Z]\hat{\mathcal{A}}_+\right)$$

$$\hat{\mathcal{D}}(\varepsilon\, z + \lambda\, \sigma) = \frac{1}{\sqrt{{}_pF_q\left([\lambda\, z + \varepsilon\, \sigma]^2\right)}}\;{}_pF_q\left([\lambda\, z + \varepsilon\, \sigma]\hat{\mathcal{A}}_+\right) \qquad (5.34)$$

so CSs with shifted argument can be defined as follows

$$|Z >= \hat{\mathcal{D}}(Z)|\,0 >= \frac{1}{\sqrt{{}_pF_q\left([Z]^2\right)}}\,{}_pF_q\left([Z]\hat{\mathcal{A}}_+\right)|\,0 >$$

$$|\,\varepsilon\, z + \lambda\,\sigma) >= \hat{\mathcal{D}}(\varepsilon\, z + \lambda\,\sigma)|\,0 >= \frac{1}{\sqrt{{}_pF_q\left([\lambda\, z + \varepsilon\,\sigma]^2\right)}}\;{}_pF_q\left([\lambda\, z + \varepsilon\,\sigma]\hat{\mathcal{A}}_+\right)|\,0 > \qquad (5.35)$$

The proof is simple.

$$|Z >= \frac{1}{\sqrt{{}_pF_q\left([Z]^2\right)}}\,{}_pF_q\left([Z]\hat{\mathcal{A}}_+\right)|\,0 >= \frac{1}{\sqrt{{}_pF_q\left([Z]^2\right)}}\sum_{n=0}^{\infty}\frac{([Z])^n}{\rho(n)}\left(\hat{\mathcal{A}}_+\right)^n|\,0 >=$$

$$= \frac{1}{\sqrt{{}_pF_q\left([Z]^2\right)}}\sum_{n=0}^{\infty}\frac{([Z])^n}{\rho(n)}\sqrt{\rho(n)}\,|\,n >= \frac{1}{\sqrt{{}_pF_q\left([Z]^2\right)}}\sum_{n=0}^{\infty}\frac{([Z])^n}{\sqrt{\rho(n)}}|\,n > \qquad (5.35)$$

respectively, in the old variables

$$|\,\varepsilon\, z + \lambda\,\sigma) >= \frac{1}{\sqrt{{}_pF_q\left([\lambda\, z + \varepsilon\,\sigma]^2\right)}}\;{}_pF_q\left([\lambda\, z + \varepsilon\,\sigma]\hat{\mathcal{A}}_+\right)|\,0 >=$$

$$= \frac{1}{\sqrt{{}_pF_q\left([\lambda\, z + \varepsilon\,\sigma]^2\right)}}\sum_{n=0}^{\infty}\frac{([\lambda\, z + \varepsilon\,\sigma])^n}{\sqrt{\rho(n)}}|\,n > \qquad (5.36)$$

In conclusion, bringing together the two ways of defining CSs with displaced argument, we will be able to write

$$|\,\varepsilon\, z + \lambda\,\sigma) >= \hat{\mathcal{D}}(\varepsilon\, z + \lambda\,\sigma)|\,0 >= \hat{\mathcal{D}}(\varepsilon\, z)\hat{\mathcal{D}}(\lambda\,\sigma)|\,0 > \qquad (5.37)$$

$$|\,\varepsilon\, z + \lambda\,\sigma) >= \frac{1}{\sqrt{{}_pF_q\left([\lambda\, z + \varepsilon\,\sigma]^2\right)}}\;{}_pF_q\left([\lambda\, z + \varepsilon\,\sigma]\hat{\mathcal{A}}_+\right)|\,0 >=$$

$$= \frac{1}{\sqrt{{}_pF_q\left(|\,\lambda\, z\,|^2\right)}}\frac{1}{\sqrt{{}_pF_q\left(|\,\varepsilon\,\sigma\,|^2\right)}}\,{}_pF_q\left(\lambda\, z\hat{\mathcal{A}}_+\right){}_pF_q\left(\varepsilon\,\sigma\hat{\mathcal{A}}_+\right)|\,0 > \qquad (5.38)$$

From here the operational equalities for generalized displacement operators follow finally, as a conclusion

$$\hat{\mathcal{D}}(\varepsilon\, z + \lambda\,\sigma) = \hat{\mathcal{D}}(\varepsilon\, z)\hat{\mathcal{D}}(\lambda\,\sigma) \qquad (5.39)$$

$$ {}_pF_q\left(\left[\lambda z+\varepsilon\sigma\right]\hat{\mathcal{A}}_+\right)=\frac{{}_pF_q\left(\left[\lambda z+\varepsilon\sigma\right]^2\right)}{\sqrt{{}_pF_q\left(|\lambda z|^2\right)}\sqrt{{}_pF_q\left(|\varepsilon\sigma|^2\right)}}\,{}_pF_q\left(\lambda z\hat{\mathcal{A}}_+\right){}_pF_q\left(\varepsilon\sigma\hat{\mathcal{A}}_+\right) \tag{5.40}$$

Obviously, let's not forget that the notation $\left[\lambda z+\varepsilon\sigma\right]$ has the meaning of Newton's generalized binomial.

For $\varepsilon=\lambda=1$ , $p=q=0$ , $\tilde{\boldsymbol{a}}=\tilde{\boldsymbol{b}}$ and considering $\sigma$ as a constant, the relation becomes

$$ \int\frac{d^2z}{\pi}G_{0,1}^{1,0}\left(|\left[z+\sigma\right]|^2\,|\,0\right){}_0F_0\left(\left[z+\sigma\right]\hat{\mathcal{A}}_+\right){}_0F_0\left(\left[z^*+\sigma\right]\hat{\mathcal{A}}_-\right)= \\ =\#{}_0F_0(\hat{\mathcal{A}}_+\hat{\mathcal{A}}_-)\#=\#\exp(\hat{\mathcal{A}}_+\hat{\mathcal{A}}_-)\# \tag{5.41}$$

and, because $\left[z+\sigma\right]=z+\sigma$, and $\rho(1)=1$, as well as of the following limit

$$ \lim_{\substack{p=q=0\\ \boldsymbol{a}=\boldsymbol{b}}}\,{}_pF_q(\tilde{\boldsymbol{a}}\,;\tilde{\boldsymbol{b}}\,;x)={}_0F_0(/\,;/\,;x)=\exp\left(x\right) \tag{5.42}$$

$$ \lim_{\substack{p=q=0\\ \boldsymbol{a}=\boldsymbol{b}}}G_{p,\,q+1}^{q+1,0}\left(x\left|\begin{matrix} /\,; & \tilde{\boldsymbol{a}}\,\boldsymbol{-1}; \\ 0\,,\tilde{\boldsymbol{b}}\,\boldsymbol{-1}\,; & / \end{matrix}\right.\right)=G_{0,1}^{1,0}\left(|z|^2\,|\,0\right)=\exp\left(-x\right) \tag{5.43}$$

we will have

$$ \int\frac{d^2z}{\pi}e^{-|z+\sigma|^2}\#e^{(z+\sigma)\hat{\mathcal{A}}_+}e^{\left(z^*+\sigma^*\right)\hat{\mathcal{A}}_-}\#=\#e^{\hat{\mathcal{A}}_+\hat{\mathcal{A}}_-}\# \tag{5.44}$$

according to [4].

This integral can be transformed as follows

$$ \int\frac{d^2z}{\pi}e^{-|z|^2+\left(\hat{\mathcal{A}}_+-\sigma^*\right)z+\left(\hat{\mathcal{A}}_--\sigma\right)z^*}\#=\#e^{\left(\hat{\mathcal{A}}_+-\sigma^*\right)\left(\hat{\mathcal{A}}_--\sigma\right)}\# \tag{5.45}$$

## 6. Particular cases

Let us specify the values of the real constants $\varepsilon$ and $\lambda$ .

***a) Case*** $\varepsilon=1$ ***and*** $\lambda=0$ ***(generalized CSs with non-displaced argument).*** This is the usual case of obtaining expressions for generalized CSs. Since the variable is no longer a binomial, we can give up the square brace $\left[...\right]$, so $\left[z\right]=z$ , and the expansion of the generalized CSs is then

$$ |\,z>=\frac{1}{\sqrt{{}_pF_q(|\,z\,|^2)}}\sum_{n=0}^{\infty}\frac{z^n}{\sqrt{\rho(n)}}|\,n>=\frac{1}{\sqrt{{}_pF_q(|\,z\,|^2)}}\,{}_pF_q(z\hat{\mathcal{A}}_+)|\,0> \tag{6.01}$$

with the generalized analogue of the displacement operator

$$\hat{\mathfrak{D}}(z) = \frac{1}{\sqrt{{}_pF_q(|z|^2)}}\ {}_pF_q(z\hat{\mathcal{A}}_+)\ {}_pF_q(z^*\hat{\mathcal{A}}_-) \tag{6.02}$$

The corresponding expression of the integration measure

$$d\mu(z) = \Gamma(a/b)\, d\left(|z|^2\right)\frac{d\varphi_z}{2\pi} G_{p,\,q+1}^{q+1,0}\left(|z|^2 \left| \begin{array}{cc} /\ ; & \{a_i-1\}_1^p \\ 0\,,\{b_j-1\}_1^q\ ; & / \end{array} \right.\right) {}_pF_q\left(|z|^2\right) \tag{6.03}$$

guarantees the validity of the decomposition relation of the unit operator, so that the corresponding integral of the Stieltjes moments is

$$\int_0^\infty d\left(|z|^2\right) G_{p,\,q+1}^{q+1,0}\left(|z|^2 \left| \begin{array}{cc} /\ ; & \{a_i-1\}_1^p \\ 0\,,\{b_j-1\}_1^q\ ; & / \end{array} \right.\right)\left(|z|^2\right)^l = \frac{1}{\Gamma\left(\tilde{\boldsymbol{a}}/\tilde{\boldsymbol{b}}\right)}\rho(l) = l!\frac{\prod_{j=1}^{q}\Gamma\left(b_j+l\right)}{\prod_{i=1}^{p}\Gamma\left(a_i+l\right)} \tag{6.04}$$

Let's highlight the following two subcases, which can also be considered as sequels to the previous case.

***a1) Subcase*** $\varepsilon = 1$ **,** $\lambda = 0$**,** $p = q$**, and** $\tilde{\boldsymbol{a}} = \tilde{\boldsymbol{b}}$ ***(canonical CSs with non-displaced argument).*** The canonical CSs are

$$|z> = e^{-\frac{1}{2}|z|^2}\sum_{n=0}^{\infty}\frac{z^n}{\sqrt{n!}}|n> = e^{-\frac{1}{2}|z|^2}e^{z\hat{a}^+}|0> \quad , \tag{6.05}$$

the integration measure reducing to $d\mu(z) = \frac{d^2z}{2\pi} = d\left(|z|^2\right)\frac{d\varphi_z}{2\pi}$, and the displacement operator is the usual one

$$\hat{\mathfrak{D}}(z) = \frac{1}{\sqrt{{}_0F_0(|z|^2)}}\ {}_0F_0(z\hat{a}^+)\ {}_0F_0(-z^*\hat{a}) = e^{-\frac{1}{2}|z|^2}e^{z\hat{a}^+ - z^*\hat{a}} \tag{6.06}$$

By customizing the indices $p$ and $q$, as well as the sets of coefficients $\tilde{\boldsymbol{a}}$ and $\tilde{\boldsymbol{b}}$, numerous examples can be obtained, as well as applications, for the various quantum systems. We continue to reproduce only the case of the pseudoharmonic oscillator (PHO).

*a2)* ***Subcase*** $\varepsilon = 1$ **,** $\lambda = 0$**,** $p = 0$**,** $q = 1$ **,** $\tilde{\boldsymbol{a}} = /$ , $\tilde{\boldsymbol{b}} = \varsigma + \frac{1}{2}$ ***(corresponding to the pseudoharmonic oscillator, PHO)***

The pseudoharmonic oscillator (PHO) is an exactly solvable quantum model, proposed by Goldman and Kryvchenkov [19] and their radial part of potential in radial coordinate $r$ is

$$V(r) = \frac{1}{2}m\omega^2r^2 + \frac{\hbar^2}{2m}\varsigma(\varsigma-1)\frac{1}{r^2} \tag{6.07}$$

where $m$ , $\omega$ and $\varsigma$ represent respectively: the mass of particle (or the reduced mass if this potential describe the nuclear motion of diatomic molecule), the frequency, and the strength of the external field.

The corresponding creation / annihilation operators for the PHO are

$$\hat{\boldsymbol{J}}_{+} | n; \varsigma >= \sqrt{(n+1)\left(n+\varsigma+\frac{1}{2}\right)} | n+1; \varsigma > \ , \quad \hat{\boldsymbol{J}}_{-} | n; \varsigma >= \sqrt{n\left(n+\varsigma-\frac{1}{2}\right)} | n-1; \varsigma > \quad (6.08)$$

and the generalized CSs are defined as

$$\hat{\boldsymbol{J}}_{-} | z; \varsigma >= z | z; \varsigma > \quad (6.09)$$

and are expanded in the Fock-vectors basis as follows:

$$\begin{aligned} | z; \varsigma > &= \frac{1}{\sqrt{{}_0F_1(/\ ; \varsigma+\frac{1}{2} ; |z|^2)}} \sum_{n=0}^{\infty} \frac{z^n}{\sqrt{n!\left(\varsigma+\frac{1}{2}\right)_n}} | n; \varsigma > = \\ &= \frac{1}{\sqrt{{}_0F_1(/\ ; \varsigma+\frac{1}{2} ; |z|^2)}} {}_0F_1(/\ ; \varsigma+\frac{1}{2} ; z\hat{\boldsymbol{J}}_{+}) | 0; \varsigma > \end{aligned} \quad (6.10)$$

The normalization hypergeometric function is connected with the modified Bessel function of the first kind

$${}_0F_1(/\ ; \varsigma+\frac{1}{2} ; |z|^2) = \Gamma\left(\varsigma+\frac{1}{2}\right) |z|^{\frac{1}{2}-\varsigma} I_{\varsigma-\frac{1}{2}}(2|z|) \quad (6.11)$$

so that the CSs becomes [20]

$$| z; \varsigma >= \sqrt{\frac{|z|^{\varsigma-\frac{1}{2}}}{I_{\varsigma-\frac{1}{2}}(2|z|)}} \sum_{n=0}^{\infty} \frac{z^n}{\sqrt{n!\,\Gamma\left(\varsigma+\frac{1}{2}+n\right)}} | n; \varsigma > \quad (6.11)$$

Then, the corresponding generalized analogue of the displaced operator acting on the vacuum ket state of the PHO, $|0; \varsigma >$ is

$$\hat{\mathfrak{D}}(z) \equiv \frac{1}{\sqrt{{}_0F_1(/\ ; \varsigma+\frac{1}{2} ; |z|^2)}} {}_0F_1(/\ ; \varsigma+\frac{1}{2} ; z\hat{\boldsymbol{J}}_{+})\ {}_0F_1(/\ ; \varsigma+\frac{1}{2} ; -z^*\hat{\boldsymbol{J}}_{-}) \ , \quad (6.12)$$

***b) Case $\varepsilon = 1$ and $\lambda = 1$ (CSs with argument sum of two complex numbers)***

The expression for generalized CSs with shifted argument is obtained from Eq. (4.13).

$$| z+\sigma >= \frac{1}{\sqrt{{}_pF_q(|[z+\sigma]|^2)}} \sum_{n=0}^{\infty} \frac{[z+\sigma]^n}{\sqrt{\rho(n)}} | n >= \frac{1}{\sqrt{{}_pF_q(|[z+\sigma]|^2)}} {}_pF_q\left([z+\sigma]\hat{\boldsymbol{\mathcal{A}}}_{+}\right) | 0 > \quad (6.13)$$

and similar their conjugate counterpart.

As we stated previously, the expression in the square parenthesis $[...]$ can be interpreted as a *generalized binomial sum*.

$$[z+\sigma]^{l} \equiv \sum_{n=0}^{l} \binom{\rho(l)}{\rho(n)} z^{l-n} \sigma^n \quad (6.14)$$

Using the DOOT rules, we can write CSs propagator

$$|\, z+\sigma><z+\sigma| = \frac{1}{\#_p F_q(\hat{\mathcal{A}}_+\hat{\mathcal{A}}_-)\#}\frac{1}{{}_pF_q(|[z+\sigma]|^2)}\#_p F_q([z+\sigma]\hat{\mathcal{A}}_+)\ {}_pF_q([z^*+\sigma^*]\hat{\mathcal{A}}_-)\# \quad (6.15)$$

To check if the normalization function ${}_pF_q(|[z+\sigma]|^2)$ is correct, let's construct the scalar product of CSs with itself

$$\begin{aligned} <z+\sigma\,|\,z+\sigma> &= \frac{1}{{}_pF_q(|[z+\sigma]|^2)}\sum_{l=0}^{\infty}\frac{[z+\sigma]^l}{\sqrt{\rho(l)}}\sum_{s=0}^{\infty}\frac{[z^*+\sigma^*]^s}{\sqrt{\rho(s)}}<s\,|\,l> = \\ &= \frac{1}{{}_pF_q(|[z+\sigma]|^2)}\sum_{l=0}^{\infty}\frac{(|[z+\sigma]|^2)^l}{\rho(l)} = 1 \end{aligned} \quad (6.16)$$

from which, using the definition of generalized hypergeometric function, the above equality is obtained.

The fundamental property of CSs, i.e. the decomposition relation of the unity operator is also satisfied. Since the variables $z$ and $\sigma$ have "equal rights", the expression must be symmetrical:

$$\iint d\mu(z,\sigma)\,|\,z+\sigma><z+\sigma\,| = 1 \quad (6.17)$$

Therefore, we will look for the integration measure in the form

$$\begin{aligned} d\mu(z,\sigma) &= \frac{d^2z}{\pi}\frac{d^2\sigma}{\pi}h_z(|\,z\,|)h_\sigma(|\,\sigma\,|)_pF_q(|[z+\sigma]|^2) = \\ &= d(|\,z\,|^2)\frac{d\varphi_z}{2\pi}h_z(|\,z\,|)d(|\,\sigma\,|^2)\frac{d\varphi_\sigma}{2\pi}h_\sigma(|\,\sigma\,|)_pF_q(|[z+\sigma]|^2) \end{aligned} \quad (6.18)$$

and we need to find the weighting functions $h_z(|\,z\,|)$ and $h_\sigma(|\,\sigma\,|)$.

So we will have

$$\int\frac{d^2z}{\pi}h_z(|\,z\,|)\int\frac{d^2\sigma}{\pi}h_\sigma(|\,\sigma\,|)\sum_{l=0}^{\infty}\frac{[z+\sigma]^l}{\sqrt{\rho(l)}}|\,l>\sum_{s=0}^{\infty}\frac{[z^*+\sigma^*]^s}{\sqrt{\rho(s)}}<s\,| = 1 \quad (6.19)$$

$$\sum_{l=0}^{\infty}\frac{|\,l>}{\sqrt{\rho(l)}}\sum_{s=0}^{\infty}\frac{<s\,|}{\sqrt{\rho(s)}}\int\frac{d^2z}{\pi}h_z(|\,z\,|)\int\frac{d^2\sigma}{\pi}h_\sigma(|\,\sigma\,|)\,[z+\sigma]^l[z^*+\sigma^*]^s = 1 \quad (6.20)$$

Let's break down the product below

$$[z+\sigma]^l[z^*+\sigma^*]^s = \sum_{n=0}^{l}\begin{pmatrix}\rho(l)\\ \rho(n)\end{pmatrix}z^{l-n}\sigma^n\sum_{m=0}^{s}\begin{pmatrix}\rho(s)\\ \rho(m)\end{pmatrix}(z^*)^{s-m}(\sigma^*)^m \quad (6.21)$$

In order to be able to satisfy the closing relation of the Fock vectors

$$\sum_{l=0}^{\infty}|\,l><l\,| = 1 \quad (6.22)$$

it is obvious, from now on, that we must have equality between the summary indices $s=l$.

Consequently, we will write

$$\sum_{l=0}^{\infty}\frac{|l><l|}{\rho(l)}\sum_{n=0}^{l}\binom{\rho(l)}{\rho(n)}\int\frac{d^2z}{\pi}h_z(|z|)z^{l-n}\left(z^*\right)^{l-m}\sum_{m=0}^{l}\binom{\rho(s)}{\rho(m)}\int\frac{d^2\sigma}{\pi}h_\sigma(|\sigma|)\sigma^n\left(\sigma^*\right)^m=1 \quad (6.23)$$

The angular integrals above will be different from zero only if the powers of the complex variable and the corresponding complex conjugate are equal. For example:

$$\int\frac{d^2\sigma}{\pi}h_\sigma(|\sigma|)\sigma^n\left(\sigma^*\right)^m=\int_0^\infty d\left(|\sigma|^2\right)h_\sigma(|\sigma|)\int_0^{2\pi}\frac{d\varphi_\sigma}{2\pi}\sigma^n\left(\sigma^*\right)^m=$$
$$=\int_0^\infty d(|\sigma|)h_\sigma(|\sigma|)\left(|\sigma|^2\right)^n\delta_{nm}=\begin{cases}0, & n\neq m\\ \neq 0, & n=m\end{cases} \quad (6.24)$$

Consequently, the relationship becomes

$$\sum_{l=0}^{\infty}|l><l|\,[g(l)]!\sum_{n=0}^{l}\binom{l}{n}\left\{\frac{1}{(l-n)!([g(l-n)]!)^2}\mathrm{Int}_z\right\}\left\{\frac{1}{n!([g(n)]!)^2}\mathrm{Int}_\sigma\right\}=1 \quad (6.25)$$

where by $\mathrm{Int}_x$ we denoted the abbreviated integrals above. For reasons of symmetry, we require that they have the following values:

$$\mathrm{Int}_{|z|}\equiv\int_0^\infty d\left(|z|^2\right)h_z(|z|)\left(|z|^2\right)^{l-n}=\frac{1}{\sqrt{2^l\,[g(l)]!}}(l-n)!([g(l-n)]!)^2 \quad (6.26)$$

$$\mathrm{Int}_{|\sigma|}\equiv\int_0^\infty d\left(|z|^2\right)h_\sigma(|\sigma|)\left(|\sigma|^2\right)^n=\frac{1}{\sqrt{2^l\,[g(l)]!}}n!([g(n)]!)^2 \quad (6.27)$$

This kind of integral is solved by first changing the exponent of the integration variable, so that it will be $s-1$, with $s$ - complex number:

$$\mathrm{Int}_{|z|}\equiv\int_0^\infty d\left(|z|^2\right)h_z(|z|)\left(|z|^2\right)^{s-1}\Bigg|_{n=l+1-s}=\frac{1}{\sqrt{2^l\,[g(l)]!}}[\Gamma(a/b)]^2\frac{\Gamma(s)\left[\prod_{j=1}^{q}\Gamma\left(b_j-1+s\right)\right]^2}{\left[\prod_{i=1}^{p}\Gamma\left(a_i-1+s\right)\right]^2} \quad (6.28)$$

$$\mathrm{Int}_{|\sigma|}\equiv\int_0^\infty d\left(|\sigma|^2\right)h_\sigma(|\sigma|)\left(|\sigma|^2\right)^{s-1}\Bigg|_{n=s-1}=\frac{1}{\sqrt{2^l\,[g(l)]!}}[\Gamma(a/b)]^2\frac{\Gamma(s)\left[\prod_{j=1}^{q}\Gamma\left(b_j-1+s\right)\right]^2}{\left[\prod_{i=1}^{p}\Gamma\left(a_i-1+s\right)\right]^2} \quad (6.29)$$

The above indexes were changed in order to be able to use the classical integral of Meijer's G functions [14]:

$$\int_0^\infty dx\, x^{s-1}\, G_{p,q}^{m,n}\left(\beta x \left| \begin{matrix} \{a_i\}_1^n\ ; & \{a_i\}_{n+1}^p \\ \{b_j\}_1^m\ ; & \{b_j\}_{m+1}^q \end{matrix} \right.\right) = \frac{1}{\beta^s} \frac{\prod_{j=1}^m \Gamma(b_j+s) \prod_{i=1}^n \Gamma(1-a_i-s)}{\prod_{j=m+1}^q \Gamma(1-b_j-s) \prod_{i=n+1}^p \Gamma(a_i+s)} \tag{6.30}$$

As a consequence, the following expressions of the weighting functions are obtained

$$h_z(|\,z\,|) = \frac{1}{\sqrt{2^l\,[g(l)]!}} [\Gamma(a/b)]^2 G_{2p,\,2q+1}^{2q+1,\,0}\left(|\,z\,|^2 \left| \begin{matrix} / & ; & \{a_i-1\}_1^p\,,\,\{a_i-1\}_1^p \\ 0\,,\ \{b_j-1\}_1^q\,,\,\{b_j-1\}_1^q & ; & / \end{matrix} \right.\right) \tag{6.31}$$

$$h_z(|\,\sigma\,|) = \frac{1}{\sqrt{2^l\,[g(l)]!}} [\Gamma(a/b)]^2 G_{2p,\,2q+1}^{2q+1,\,0}\left(|\,\sigma\,|^2 \left| \begin{matrix} / & ; & \{a_i-1\}_1^p\,,\,\{a_i-1\}_1^p \\ 0\,,\ \{b_j-1\}_1^q\,,\,\{b_j-1\}_1^q & ; & / \end{matrix} \right.\right) \tag{6.32}$$

A beautiful symmetry is observed relative to the variables $z$ and $\sigma$, which was to be expected.

Substituting these results in the decomposition relation of the unit operator and taking into account the equality

$$\sum_{m=0}^{l} \binom{l}{m} = 2^l \tag{6.33}$$

we will find that the closure relation of the Fock vectors is satisfied.

Let us now examine the particular case when $p=q=0\,,\{a_i\}_1^p=\{b_j\}_1^q$ and, consequently, $\rho(x)=x!$, $[g(0)]!=1$ which case corresponds to the canonical CSs (characteristics of the HO-1D one-dimensional harmonic oscillator).

These are, keeping in mind that for this case ${}_pF_q(\tilde{\boldsymbol{a}}\,;\tilde{\boldsymbol{b}}\,;x)={}_0F_0\ ;\ ;x)=\exp(x)$, and $[z+\sigma]^l=(z+\sigma)^l$:

$$|\,z+\sigma>=e^{-\frac{1}{2}|z+\sigma|^2}\sum_{l=0}^{\infty}\frac{(z+\sigma)^l}{\sqrt{l!}}|\,l>=e^{-\frac{1}{2}|z+\sigma|^2+(z+\sigma)A_+}\,|\,0> \tag{6.34}$$

The integration measures is

$$d\mu(z,\sigma)=\frac{d^2z}{\pi}\frac{d^2\sigma}{\pi}h_z(|\,z\,|)h_\sigma(|\,\sigma\,|)e^{|z+\sigma|^2} \tag{6.35}$$

and the weighting functions are, respectively

$$h_z(|\,z\,|)=G_{0,\,1}^{1,\,0}\left(|\,z\,|^2\,\middle|\,0\right)=e^{-|z|^2}\ ,\quad h_\sigma(|\,\sigma\,|)=G_{0,\,1}^{1,\,0}\left(|\,\sigma\,|^2\,\middle|\,0\right)=e^{-|\sigma|^2} \tag{6.36}$$

The decomposition relation of the unity operator is written, therefore

$$\iint d\mu(z,\sigma)\,|\,z+\sigma><z+\sigma\,|=\sum_{l=0}^{\infty}|\,l><l\,|\ \sum_{n=0}^{l}\binom{l}{n}\left\{\frac{1}{(l-n)!}\,\mathrm{Int}_z\right\}\left\{\frac{1}{n!}\,\mathrm{Int}_\sigma\right\}=1 \tag{6.37}$$

$$\mathrm{Int}_{|z|} = \frac{1}{\sqrt{2^l}}(l-n)! \qquad , \qquad \mathrm{Int}_{|\sigma|} = \frac{1}{\sqrt{2^l}} n! \tag{6.38}$$

from which it follows that the equality is satisfied.

## 7. Concluding remarks

In the paper we obtain the expression of CSs with shifted or displaced argument, whose argument is a linear combination of two different complex variables. In this sense, we used a pair of generalized ladder operators, $\hat{\mathcal{A}}_{-} = \hat{a}\, f(\hat{\mathcal{N}})$ and $\hat{\mathcal{A}}_{+} = f(\hat{\mathcal{N}})\,\hat{a}^{+}$ , where $f(\hat{\mathcal{N}})$ is the nonlinearity function $f(\hat{\mathcal{N}})$. These operators generate different sets of nonlinear CSs. These CSs with shifted argument were obtained in two equivalent ways: 1.) the classical method, also used in the case of canonical operators, i.e. by successively applying two displacement operators, each depending on only one complex variable; 2.) generalized method, by using the generalized displacement operators, which were defined by Mojaveri and Dehghani [16], that is, as normalized generalized hypergeometric function, which had $z\hat{\mathcal{A}}_{+}$ as its argument. In both methods, CSs are obtained by applying the normed displacement operators on the vacuum state ket $|0>$. The general results obtained were verified for several particular cases: the case of generalized CSs with non-displaced argument, with subcases - canonical CSs with non-displaced argument and CSs corresponding to the pseudoharmonic oscillator, PHO, as well as for the case of CSs with argument sum of two complex numbers. This last case is useful for obtaining the expression for the Wigner operator in quantum optics. In all calculations we used our diagonal operator ordering technique (DOOT), which is a generalization of the integration of an ordered product of operators (IWOP) technique developed by Hong-Yi Fan and applied to canonical operators. The DOOT technique is applicable to linear operators $\hat{\mathcal{A}}_{-} = \hat{a}\, f(\hat{\mathcal{N}})$ and $\hat{\mathcal{A}}_{+} = f(\hat{\mathcal{N}})\,\hat{a}^{+}$ and assumes that they commute.